\documentclass[reprint,superscriptaddress,amsmath,amssymb,aps]{revtex4-1}

\usepackage{graphicx}
\usepackage{dcolumn}
\usepackage{bm}
\usepackage{subfigure}
\usepackage{natbib} 
\usepackage{multirow}
\usepackage{soul}
\usepackage{float}
\usepackage[normalem]{ulem}
\usepackage[usenames,dvipsnames]{color}
\usepackage{appendix}
\usepackage{mathrsfs}

\newcommand{\bra}[1] { \langle #1 | }
\newcommand{\ket}[1] { | #1 \rangle }
\newcommand{\braket}[2] { \langle #1 | #2 \rangle }
\newcommand{\Bra}[1] { \Bigl\langle #1 \Bigr| }
\newcommand{\Ket}[1] { \Bigl| #1 \Bigr\rangle }
\newcommand{\Braket}[2] { \Bigl\langle #1 \Big| #2 \Bigr\rangle }

\newcommand{\displ}[0] { \mathbf{R} }
\newcommand{\ooverlap}[0] { \mathsf{O} }
\newcommand{\Umatrix}[0] { \mathsf{U} }
\newcommand{\ttt}[0] { t }
\newcommand{\Odiag}[0] {{\mathsf{Z}} }
\newcommand{\DinvO}[0] {\mathsf{W} }
\newcommand{\InvSqrtO}[0] {\mathsf{O}^{-\frac{1}{2}} }
\newcommand{\eigenval}[0] {z}
\newcommand{\unit}[0] {\mathsf{1} }
\newcommand{\Amatrix}[0] {\mathsf{A} }

\def\QE{\textsc{Quantum ESPRESSO}\,}

\newcommand{\editor}[2]{%
  \expandafter\newcommand\csname #1note\endcsname[1]{%
    \textcolor{#2}{(\textbf{#1:} ##1)}}%
  \expandafter\newcommand\csname #1\endcsname[1]{%
    \textcolor{#2}{##1}}%
  \expandafter\newcommand\csname #1cancel\endcsname[1]{%
    \textcolor{#2}{\sout{##1}}}%
  \expandafter\newcommand\csname #1change\endcsname[2]{%
    \textcolor{#2}{\sout{##1} ##2}}%
  \newenvironment{#1text}{\color{#2}}{\color{black}}
}

\begin{document}

\title{Pulay forces in density-functional theory with extended Hubbard functionals:\\ From nonorthogonalized to orthogonalized manifolds}

\author{Iurii Timrov}\email[e-mail:]{ iurii.timrov@epfl.ch}
\affiliation{Theory and Simulation of Materials (THEOS), and National Centre for Computational Design and Discovery of Novel Materials (MARVEL), \'Ecole Polytechnique F\'ed\'erale de Lausanne (EPFL), CH-1015 Lausanne, Switzerland}

\author{Francesco Aquilante}
\affiliation{Theory and Simulation of Materials (THEOS), and National Centre for Computational Design and Discovery of Novel Materials (MARVEL), \'Ecole Polytechnique F\'ed\'erale de Lausanne (EPFL), CH-1015 Lausanne, Switzerland}

\author{Luca Binci}
\affiliation{Theory and Simulation of Materials (THEOS), and National Centre for Computational Design and Discovery of Novel Materials (MARVEL), \'Ecole Polytechnique F\'ed\'erale de Lausanne (EPFL), CH-1015 Lausanne, Switzerland}

\author{Matteo Cococcioni}
\affiliation{Department of Physics, University of Pavia, I-27100 Pavia, Italy}

\author{Nicola Marzari}
\affiliation{Theory and Simulation of Materials (THEOS), and National Centre for Computational Design and Discovery of Novel Materials (MARVEL), \'Ecole Polytechnique F\'ed\'erale de Lausanne (EPFL), CH-1015 Lausanne, Switzerland}

\date{\today}

\begin{abstract}
We present a derivation of the exact expression for Pulay forces in density-functional theory calculations augmented with extended Hubbard functionals, and arising from the use of orthogonalized atomic orbitals as projectors for the Hubbard manifold. The derivative of the inverse square root of the orbital overlap matrix is obtained as a closed-form solution of the associated Lyapunov (Sylvester) equation. The expression for the resulting contribution to the forces is presented in the framework of ultrasoft pseudopotentials and the projector-augmented-wave method, and using a plane wave basis set. We have benchmarked the present implementation with respect to finite differences of total energies for the case of NiO, finding excellent agreement. Owing to the accuracy of Hubbard-corrected density-functional theory calculations -- provided the Hubbard parameters are computed for the manifold under consideration -- the present work paves the way for systematic studies of solid-state and molecular transition-metal and rare-earth compounds. 
\end{abstract}

\maketitle

\section{Introduction}
\label{sec:Introduction}

Density-functional theory (DFT)~\cite{Hohenberg:1964,Kohn:1965} with approximate exchange-correlation functionals has been remarkably successful in predicting ground-state properties of a large variety of systems. However, most functionals [such as the local-density approximation (LDA) and generalized-gradient approximation (GGA)] fail in capturing both qualitatively and quantitatively the ground state of systems with strongly localized electrons (typically, of $d$ and/or $f$ character), due to large self-interaction errors (SIE)~\cite{Perdew:1981, MoriSanchez:2006}. There are different schemes that can be used in DFT to alleviate SIE; in particular, we mention here self-interaction corrections (SIC)~\cite{Perdew:1981, Svane:1990, Vogel:1996, Filippetti:2003}, hybrid functionals~\cite{moreira:2002, cora:2004, feng:2004, Alfredsson:2004, Tran:2006, Chevrier:2010, Seo:2015}, meta-GGA functionals~\cite{Tao:2003, Perdew:2009, Sun:2012, Zhao:2006, delCampo:2012, Sun:2015}, and DFT+$U$~\cite{anisimov:1991, Liechtenstein:1995, dudarev:1998, Kulik:2006, Kulik:2008, Kulik:2011, Himmetoglu:2014} or its extension DFT+$U$+$V$~\cite{Campo:2010, TancogneDejean:2020, Lee:2020}.

DFT with (extended) Hubbard functionals -- DFT+$U$ (DFT+$U$+$V$) -- is popular due to its simplicity and much improved accuracy in describing structural, electronic, and magnetic ground-state properties of transition-metal and rare-earth compounds~\cite{Anisimov:2010, Himmetoglu:2014} by removing self-interactions for a subset of electronic states (i.e. states in the Hubbard manifold)~\cite{Kulik:2006}. While DFT+$U$ with an on-site Hubbard $U$ correction is effective for many systems with strongly localized electrons, the inter-site Hubbard $V$ contribution is crucial in many systems having also a strong covalent hybridization for the same orbitals; case studies have been the evaluation of voltages in Li-ion batteries~\cite{Cococcioni:2019}, the determination of formation energies of oxygen vacancies in perovskites~\cite{Ricca:2020}, and geometries and energetics in molecular systems~\cite{Kulik:2011}. Key aspects of these methods are: $i)$~the choice of the Hubbard parameters (on-site $U$ and inter-site $V$), and $ii)$~the choice of projector functions that are used to construct the Hubbard manifold. These two fundamental points are not independent. Hubbard parameters that are computed from first principles (e.g. using constrained DFT (cDFT)~\cite{Dederichs:1984, Mcmahan:1988, Gunnarsson:1989, Hybertsen:1989, Gunnarsson:1990, Pickett:1998, Solovyev:2005, Nakamura:2006, Shishkin:2016, Nawa:2018}, Hartree-Fock-based approaches~\cite{Mosey:2007, Mosey:2008, Andriotis:2010, Agapito:2015, TancogneDejean:2020}, constrained random-phase approximation (cRPA)~\cite{Springer:1998, Kotani:2000, Aryasetiawan:2004, Aryasetiawan:2006, Sasioglu:2011, Vaugier:2012, Amadon:2014, Seth:2017, Panda:2017, Nakamura:2020}, or linear-response theory (LRT)~\cite{Cococcioni:2005, Timrov:2018}) vary substantially, depending on the choice of the projector functions for the Hubbard manifold, pseudopotentials and the oxidation state~\cite{Kulik:2008, Shishkin:2016}, exchange-correlation functionals, and chemical environment of the Hubbard atoms~\cite{Floris:2020, Bennett:2019}. Therefore, it is crucial to use the $U$ and $V$ parameters consistently with the Hubbard manifold and other technicalities (pseudopotentials, oxidation states, functionals, etc.) which were used to compute them. In addition, it is important to stress that the final quantities of interest (energies, bond lengths, etc.) are 
insensitive to large variations (2-3~eV) in interaction parameters computed using pseudopotentials generated in different oxidation states provided that these interaction parameters are computed self-consistently (e.g. using LRT) (see the Appendix of Ref.~\cite{Kulik:2008}).

Since the early days~\cite{anisimov:1991}, DFT+$U$ became widely used and implemented in different electronic-structure codes. However, different projector functions for the Hubbard manifold are used in the available implementations (see below), which makes it difficult to compare results obtained with different codes; in addition, it is still a common practice to choose empirical values of $U$, disregarding the underlying definition of the Hubbard manifold. In Ref.~\cite{Wang:2016}, an effort was made to compare and analyze various types of projector functions for a set of $U$ values, and quite large variations in the results (e.g. density of states, energy differences) were obtained, especially for systems with strong covalent interactions compared to the ones with ionic interactions.

There are quite many possible projector functions to use as a basis for the Hubbard manifold (see e.g. Ref.~\cite{Tablero:2008}). In particular, we highlight here nonorthogonalized atomic orbitals (NAO)~\cite{Cococcioni:2005, Amadon:2008}, orthogonalized atomic orbitals (OAO)~\cite{Cococcioni:2019, Ricca:2020, Timrov:2020}, nonorthogonalized Wannier functions (NWF)~\cite{ORegan:2010}, orthogonalized Wannier functions (OWF)~\cite{Korotin:2012}, linearized augmented plane-wave (LAPW) approaches~\cite{Shick:1999}, and projector-augmented-wave (PAW) projector functions~\cite{Bengone:2000, Rohrbach:2003}. A common feature of all these projector functions is that they are spatially localized and depend explicitly on atomic positions; hence, an extra term appears when computing derivatives of the Hubbard corrective energy with respect to atomic displacements (Pulay force) or strain (Pulay stress). The expressions for Pulay forces in the context of Hubbard-corrected DFT were derived for NAO~\cite{Cococcioni:2010, Cococcioni:2012}, NWF~\cite{Roychoudhury:2018}, OWF~\cite{Novoselov:2015}, and  LAPW~\cite{Tran:2008}, but no derivation was made so far for the case of OAO. The scope of the present work is to extend the existing expression for Pulay forces using NAO~\cite{Cococcioni:2010, Cococcioni:2012} to orthogonalized basis sets based on the L\"owdin scheme~\cite{Lowdin:1950, Mayer:2002}.

The importance of this development is that NAO and OAO are of special interest due to the simplicity of their implementation and transparency in their practical use. In particular, DFT+$U$ calculations with NAO have proven to be accurate for simulations of various properties in many materials~\cite{Zhou:2004, Hsu:2009, Ricca:2019, Floris:2020, Sun:2020}. However, NAO have also drawbacks dictated by the fact that atomic orbitals can have long tails that extend to a significant spatial range, and as a consequence the formal occupation numbers for these orbitals can be substantially overestimated~\cite{Wang:2016}. In some NAO-based works~\cite{Amadon:2008} these tails are truncated (i.e. atomic orbitals are zero outside of ``atomic spheres''), and even in LAPW and PAW-based approaches there is some ambiguity in the definition of the cutoff radius at which the projector functions are truncated~\cite{Nawa:2018}. Such ambiguity has implications on the final quantities of interest that are computed~\cite{Wang:2016, Nawa:2018}; in addition, in the case of NAO with long tails, the Hubbard correction is essentially applied twice in the overlap regions between atoms, which can further enhance spurious unphysical effects in Hubbard-corrected DFT calculations. These issues can be removed by orthogonalizing the atomic orbitals among all atoms: this insures that Hubbard corrections are applied only once to the respective Hubbard manifolds. Moreover, DFT+$U$ with OAO captures some inter-site corrections through orthogonalization of the orbitals of one atom combined with those from neighbor sites, thus making DFT+$U$ closer to the full DFT+$U$+$V$.  Therefore, OAO is a very attractive alternative to NAO, and in fact it was already observed that the former gives more accurate energetics than the latter~\cite{Cococcioni:2019, Ricca:2020}. Until the present work, though, atomic and cell relaxations with OAO were not possible due to the difficulty in evaluating the derivative of the inverse square root of the orbital overlap matrix, which appears when using the L\"owdin scheme~\cite{Lowdin:1950, Mayer:2002}.

In this work, we present a derivation that allows us to calculate Pulay (Hubbard) forces for the case of OAO projector functions, by starting from the expressions for the Hubbard force in the case of NAO~\cite{Cococcioni:2010, Cococcioni:2012} and using the Hubbard parameters $U$ and $V$ computed using density-functional perturbation theory (DFPT)~\cite{Timrov:2018, Timrov:2020b}. We present a detailed mathematical formulation of the derivative of the inverse square root of the orbital overlap matrix as a closed-form solution of the Lyapunov (Sylvester) equation, which is the main result of the present formalism, and we compare it with other techniques that were used in literature to compute such a derivative. The formalism is presented in the framework of DFT+$U$+$V$, i.e. by taking into account not only Hubbard forces coming from the on-site $U$ term but also from the inter-site $V$ term. For the sake of generality, we present the derivation in the case of ultrasoft (US) pseudopotentials (PPs) and PAW. It is worth to note that we have also derived and implemented the formalism for the Hubbard stress using OAO, but this will not be discussed here because this is not the focus of this paper.

The paper is organized as follows: Section~\ref{sec:dft_hubbard_basics} presents the basics of DFT+$U$ and DFT+$U$+$V$ in the framework of US and PAW PPs; in Sec.~\ref{sec:Hubbard_manifold} we discuss NAO and OAO; Sec.~\ref{sec:force_general} discusses the Hellmann-Feynman theorem and its generalization to US and PAW PPs; in Sec.~\ref{sec:force_atomic} and Sec.~\ref{sec:force_orthoatomic} we present the derivation of Hubbard forces in the cases of NAO and OAO, respectively; Sec.~\ref{sec:technical_details} contains technical details of our calculations; in Sec.~\ref{sec:results} we benchmark the implementation of Hubbard forces using OAO versus the finite difference method, and make comparisons of total and Hubbard forces in the case of OAO and NAO; and finally, in Sec.~\ref{sec:conclusions} we give our conclusions. In Appendix~\ref{app:Taylor} we give an alternative (approximate) expression for the derivative of the inverse square root of the overlap matrix based on the Taylor series expansion, in Appendix~\ref{app:Lyap_solution} we present a proof that the solution of the Lyapunov equation can be written in a closed form, and in Appendix~\ref{sec:Scaling} we present the discussion about the computational scaling of forces. Hartree atomic units are used throughout the paper. For the sake of simplicity, the formalism is presented for insulators. We will use notations and definitions similar to those in Ref.~\cite{Dalcorso:2001}. Quantum-mechanical operators will be indicated with a hat on top of capital letters (``$\hat{A}$''), while for matrices we will use a special font (``$\Amatrix$'').

\section{Theory}
\label{sec:Theory}

In this section we present the formalism for calculation of Hubbard forces starting from the expression for the Hubbard energy in the framework of DFT with extended Hubbard functionals. The main results of this paper can be divided into two parts: $(i)$~generalization of the DFT+$U$ formalism to the framework of DFT+$U$+$V$ and to the use of US or PAW PPs in the context of NAO, and $(ii)$~generalization of ``$(i)$'' to OAO. The former is discussed in Secs.~\ref{sec:force_general} and \ref{sec:force_atomic}, while the latter in Sec.~\ref{sec:force_orthoatomic}.

\subsection{DFT+$U$+$V$}
\label{sec:dft_hubbard_basics}

In this section we briefly recall the basics of the DFT+$U$+$V$ formalism in the simplified rotationally-invariant form, which were presented in Refs.~\cite{dudarev:1998, Cococcioni:2005, Campo:2010} for norm-conserving (NC) PPs, and later were extended to US and PAW PPs in Refs.~\cite{Floris:2020, Timrov:2020b}. This reminder is needed, because it will be our starting point for the derivation of expressions for forces.

Similarly to DFT+$U$, DFT+$U$+$V$ is also based on an additive correction to the approximate DFT energy functional~\cite{Campo:2010}:
\begin{equation}
E_\mathrm{TOT} = E_{\mathrm{DFT}} + E_\mathrm{Hub} \,,
\label{eq:Edft_plus_u}
\end{equation}
where $E_\mathrm{TOT}$ is the total energy functional, $E_{\mathrm{DFT}}$ is the standard DFT energy functional based on LDA or GGA, and $E_\mathrm{Hub}$ is the Hubbard energy functional which is the difference between the Hubbard term and its mean-field approximation which is subtracted to avoid the double-counting of interactions already included in $E_{\mathrm{DFT}}$. In the present work, this latter term is shaped according to the popular fully localized limit~\cite{anisimov:1997, dudarev:1998, Himmetoglu:2014}. The generalized Hubbard corrective energy is defined as~\cite{Campo:2010}:
\begin{eqnarray}
E_\mathrm{Hub} & = & \frac{1}{2} \sum_I \sum_{\sigma m_1 m_2} 
U^I \left( \delta_{m_1 m_2} - n^{II \sigma}_{m_1 m_2} \right) n^{II \sigma}_{m_2 m_1} \nonumber \\
& & - \frac{1}{2} \sum_{I} \sum_{J (J \ne I)}^* \sum_{\sigma m_1 m_2} V^{I J} 
n^{I J \sigma}_{m_1 m_2} n^{J I \sigma}_{m_2 m_1} \,,
\label{eq:Edftu}
\end{eqnarray}
where $I$ and $J$ are the atomic site indices, $m_1$ and $m_2$ are the magnetic quantum numbers associated with a specific angular momentum (i.e. orbitals of the Hubbard manifold) of atoms $I$ and $J$, respectively, $U^I$ and $V^{I J}$ are the on-site and inter-site Hubbard parameters, respectively, and the star in the sum on the second line denotes that for each atom $I$ index $J$ covers all its nearest neighbors up to a given distance (or belonging to a given shell). Typically, only nearest neighbors are considered in the inter-site term, but the formalism is general and hence allows to include next-nearest neighbors and even further ones when needed.

In Eq.~\eqref{eq:Edftu}, $n^{I J \sigma}_{m_1 m_2}$ are the generalized occupation matrices that are based on a projection of Kohn-Sham (KS) states $\psi_{i\sigma}(\mathbf{r})$ on the Hubbard manifold specific for each atom $\{ \varphi^{I}_{m}(\mathbf{r}) \}$: 
\begin{eqnarray}
n^{I J \sigma}_{m_1 m_2} & = & \sum_{i}
\bra{\psi_{i\sigma}} \hat{S} \ket{\varphi^{J}_{m_2}} \bra{\varphi^{I}_{m_1}} \hat{S}
\ket{\psi_{i\sigma}} \nonumber \\
& = & \sum_{i} \bra{\psi_{i\sigma}} \hat{P}^{J I}_{m_2 m_1} \ket{\psi_{i\sigma}} \,,
\label{eq:occ_matrix}
\end{eqnarray}
where
\begin{equation}
\hat{P}^{J I}_{m_2 m_1} = \hat{S} \ket{\varphi^{J}_{m_2}} \bra{\varphi^{I}_{m_1}} \hat{S} \,.
\label{eq:Pm1m2us}
\end{equation}
Here, index $i$ labels KS states, $\hat{P}^{J I}_{m_2 m_1}$ is the generalized projector on Hubbard manifolds of atoms $J$ and $I$, and $\varphi^I_{m_1}(\mathbf{r}) \equiv \varphi^{\gamma(I)}_{m_1}(\mathbf{r} - \mathbf{R}_I)$ are the functions centered on the $I$th atom of type $\gamma(I)$ at the position $\mathbf{R}_I$. Depending on the atomic type, functions $\varphi^I_{m_1}(\mathbf{r})$ can be either localized functions (of $d$ or $f$ character like for transition-metal and rare-earth elements) or spread functions (of $s$ or $p$ character for other elements). In the following sections we will discuss in more details these functions and their role in Hubbard-corrected DFT calculations. 

In Eq.~\eqref{eq:Pm1m2us}, $\hat{S}$ is the operator of the US or PAW PPs schemes, which reads
\begin{equation}
\hat{S} = 1 + \sum_{I\mu\nu} q^{\gamma(I)}_{\mu\nu} \, \ket{\beta^{I}_\mu} \bra{\beta^{I}_\nu} \,,
\label{eq:S_overlap}
\end{equation}
where
\begin{equation}
q^{\gamma(I)}_{\mu\nu} = 
\int Q^{\gamma(I)}_{\mu\nu}(\mathbf{r}) \, d\mathbf{r} \,.
\label{intq}
\end{equation}
Here, $\mu$ and $\nu$ are the indices which label $Q$ and $\beta$ functions, $Q^{\gamma(I)}_{\mu\nu}(\mathbf{r} - \mathbf{R}_I)$ are the localized augmentation functions pertaining to the pseudopotential of the $I$th atom, $\beta^{I}_\mu(\mathbf{r}) \equiv \beta^{\gamma(I)}_\mu(\mathbf{r} - \mathbf{R}_I)$ are the so-called projector functions of the US or PAW PPs schemes [not to be confused with the projector functions on the Hubbard manifold, $\varphi^I_{m_1}(\mathbf{r})$] that are localized on the $I$th atom and vanish outside spheres centered on atoms~\cite{Vanderbilt:1990}, and the integration in Eq.~\eqref{intq} is performed over a crystal volume.

In DFT+$U$, only the first line in Eq.~\eqref{eq:Edftu} is preserved, while the term on the second line is zero because the inter-site interactions are neglected, i.e. $V^{IJ}=0$. In this case, which corrects only the on-site interactions, it is useful to adopt the following notation: $n^{I\sigma}_{m_1m_2} \equiv n^{II\sigma}_{m_1m_2}$ and $\hat P^I_{m_1 m_2} \equiv \hat P^{II}_{m_1 m_2}$. It is easy to see from Eq.~\eqref{eq:Edftu} that the two terms of the corrective energy functional, proportional to the on-site ($U^{I}$) and inter-site ($V^{IJ}$) interactions, counteract each other. In fact, while the on-site term favors localization on atomic sites (by suppressing inter-site hybridization of orbitals), the inter-site terms restore and stabilize hybridized states in the interstitial regions between neighboring atoms that are characteristic for covalent interactions. Therefore, in systems with predominantly ionic interactions DFT+$U$ is expected to be sufficient and a good level of approximation, while in systems with predominantly covalent interactions DFT+$U$+$V$ is needed.

For the purpose of this work it is important to discuss the contribution to the KS potential stemming from the extended Hubbard functional [see Eq.~\eqref{eq:Edftu}]. The action of this term on KS wavefunctions can be easily obtained by taking a functional derivative of $E_{\mathrm{TOT}}$ [see Eq.~\eqref{eq:Edft_plus_u}] with respect to the complex conjugate of KS wavefunctions~\cite{Campo:2010, Timrov:Note:2018:errorJPCM}. The term corresponding to the functional derivative of $E_\mathrm{Hub}$ [see Eq.~\eqref{eq:Edftu}] reads:
\begin{eqnarray}
\hat{V}_{\mathrm{Hub},\sigma} & = & 
\sum_I \sum_{m_1 m_2} U^{I} \left( \frac{\delta_{m_1 m_2}}{2} - 
n^{I \sigma}_{m_1 m_2} \right) \hat{P}^{I}_{m_1 m_2} \nonumber \\
& & - \sum_{I} \sum_{J (J \ne I)}^* \sum_{m_1 m_2} V^{I J} 
n^{I J \sigma}_{m_1 m_2} \hat{P}^{I J}_{m_1 m_2} \,.
\label{eq:Hub_pot_0}
\end{eqnarray}
Therefore, the generalized KS equations with the Hubbard corrections can be written as
\begin{equation}
\hat{H}_{\sigma} \ket{\psi_{i\sigma}} = 
\varepsilon_{i\sigma} \hat{S} \ket{\psi_{i\sigma}} \,,
\label{eq:KSeq_GS}
\end{equation} 
where $\varepsilon_{i\sigma}$ are the KS energies, and
\begin{equation}
\hat{H}_\sigma = \hat{H}_{\mathrm{DFT},\sigma} + \hat{V}_{\mathrm{Hub},\sigma} \,,
\label{eq:H_tot_GS}
\end{equation}
with $\hat{H}_{\mathrm{DFT},\sigma}$ being the standard DFT Hamiltonian (LDA or GGA), 
and $\hat{V}_{\mathrm{Hub},\sigma}$ is the Hubbard potential given by Eq.~\eqref{eq:Hub_pot_0}. For generalized KS equations, the orthonormality condition reads:
\begin{equation}
\bra{\psi_{i\sigma}} \hat{S} \ket{\psi_{i'\sigma'}} 
= \delta_{i i'} \delta_{\sigma \sigma'} \,.
\label{eq:orthonormality}
\end{equation}
The DFT Hamiltonian $\hat{H}_{\mathrm{DFT},\sigma}$ contains usual terms~\cite{Vanderbilt:1990, Blochl:1994}), among which the Hartree and exchange-correlation potentials depend on the charge density, which in the US and PAW PPs cases reads:
\begin{eqnarray}
& &
\rho_\sigma(\mathbf{r}) = \sum_{i}
\vert\psi_{i\sigma}(\mathbf{r})\vert^2 \nonumber \\ 
& & \hspace{0.7cm} + \sum_{i} 
\sum_{I\mu\nu} Q^{\gamma(I)}_{\mu\nu}(\mathbf{r}-\mathbf{R}_I) \,
\braket{\psi_{i\sigma}}{\beta^{I}_\mu} 
\braket{\beta^{I}_\nu}{\psi_{i\sigma}} \,. \nonumber \\
& & 
\label{eq:density_0}
\end{eqnarray}
Finally, Hubbard parameters $U^I$ and $V^{IJ}$, which are needed for the formalism presented above, can be computed from first principles using e.g. linear response theory~\cite{Cococcioni:2005} with its recent reformulation based on density-functional perturbation theory~\cite{Timrov:2018, Timrov:2020b}. As was mentioned in Sec.~\ref{sec:Introduction}, values of $U^I$ and $V^{IJ}$ depend strongly on the choice of projector functions of the Hubbard manifold, $\varphi^I_{m_1}(\mathbf{r})$, as well as on the type of PPs~\cite{Shishkin:2016}, oxidation state~\cite{Kulik:2008, Bennett:2019}, functional, and chemical composition of the system~\cite{Bennett:2019, Floris:2020}.

\subsection{Choosing projector functions for the Hubbard manifold}
\label{sec:Hubbard_manifold}

One of the key aspects of the Hubbard-corrected DFT formalism is the choice of the projector functions for the Hubbard manifold. In other words, we need to choose the basis $\{ \varphi^I_{m}(\mathbf{r}) \}$ for the projector $\hat{P}^{J I}_{m_2 m_1}$ that was introduced in Sec.~\ref{sec:dft_hubbard_basics}. In Sec.~\ref{sec:Introduction} we discussed what are the popular choices in literature for $\{ \varphi^I_{m}(\mathbf{r}) \}$. In this work, we focus our discussion only on two types of projector functions, NAO and OAO. Let us comment briefly about each of them.

NAO is one of the most simple projector functions for the Hubbard manifold, which is often a reasonable choice to represent the Hubbard manifold, especially in systems with mostly ionic character of interactions. NAO are provided with pseudopotentials, and these orbitals are orthonormal within each atom (i.e. Hubbard $d$ orbitals are orthonormal to non-Hubbard $s$ and $p$ orbitals of the same atom) but not between different atoms. However, whenever covalent interactions become important, this type of projector functions is not the best choice (see the discussion in Sec.~\ref{sec:Introduction}), and inter-site orthogonalization becomes important. 

OAO are obtained by taking atomic orbitals of each atom and then orthogonalizing them to all orbitals of all atoms in the system. In this work, we will use the L\"owdin orthogonalization method~\cite{Lowdin:1950, Mayer:2002}. By doing so, we obtain a new set of orbitals, that are all orthogonalized, which now better represent hybridizations of orbitals between neighboring sites. This choice is particularly good for setting up the Hubbard manifold, because it allows us to avoid counting Hubbard corrections twice in the interstitial regions between atoms, which is especially relevant in the case of DFT+$U$+$V$.

The Hubbard-corrected DFT formalism presented in Sec.~\ref{sec:dft_hubbard_basics} is general and hence it applies both to NAO and OAO. However, the expressions for the Hubbard forces have differences, which we will detail in the following. First, we will briefly recall the formalism for Hubbard forces that are computed using NAO (Sec.~\ref{sec:force_atomic}), and, second, we will present the generalization to OAO and highlight what are the difference with the NAO case (Sec.~\ref{sec:force_orthoatomic}).

\subsection{Total forces in DFT+$U$+$V$}
\label{sec:force_general}

In this section we discuss how to evaluate Hubbard forces starting from the expression for the total energy, Eq.~\eqref{eq:Edft_plus_u}. The main idea is based on the Hellmann-Feynman theorem which states that in the case of NC PPs the derivative of the total energy with respect to some small perturbation (in this case the perturbation is the atomic displacement) equals to the expectation value of the derivative of the Hamiltonian. However, in the case of US and PAW PPs, there is a contribution coming also from the derivatives of the $\hat{S}$ operator~\cite{Dalcorso:2001, Floris:2020}. Therefore, the total force acting on the $K$th atom upon its displacement is:
\begin{widetext}
\begin{eqnarray}
    \mathbf{F}_{\mathrm{TOT}, K} = - \frac{dE_{\mathrm{TOT}}}{d\displ_K} & = & 
    -\sum_{i\sigma} \int \frac{\delta E_{\mathrm{TOT}}}{\delta \psi^*_{i\sigma}(\mathbf{r})} \frac{d\psi^*_{i\sigma}(\mathbf{r})}{d\displ_K} d\mathbf{r} -
    \sum_{i\sigma} \int \frac{\delta E_{\mathrm{TOT}}}{\delta \psi_{i\sigma}(\mathbf{r})} \frac{d\psi_{i\sigma}(\mathbf{r})}{d\displ_K} d\mathbf{r} -
    \frac{\partial E_{\mathrm{TOT}}}{\partial \displ_K} \nonumber \\
    & = & - \sum_{i\sigma} \Bigl\{ \Bra{\frac{d\psi_{i\sigma}}{d\displ_K}} \hat{H}_\sigma \Ket{\psi_{i\sigma}} + \Bra{\psi_{i\sigma}} \hat{H}_\sigma \Ket{\frac{d\psi_{i\sigma}}{d\displ_K}} + \Bra{\psi_{i\sigma}} \frac{\partial \hat{H}_\sigma}{\partial \displ_K} \Ket{\psi_{i\sigma}} \Bigr\} \nonumber \\
    & = & - \sum_{i\sigma} \Bra{\psi_{i\sigma}} \Bigl( \frac{\partial \hat{H}_{\mathrm{DFT,\sigma}}}{\partial \displ_K} - \varepsilon_{i\sigma} \frac{\partial \hat{S}}{\partial \displ_K} \Bigr) \Ket{\psi_{i\sigma}} - \sum_{i\sigma} \Bra{\psi_{i\sigma}} \frac{\partial \hat{V}_{\mathrm{Hub},\sigma}}{\partial \displ_K} \Ket{\psi_{i\sigma}} \,,
    \label{eq:HF_theorem}
\end{eqnarray}
\end{widetext}
where the first two terms (in the last row) come from the standard DFT in the US or PAW PPs formalism, and the last term is the Hubbard force:
\begin{equation}
    \mathbf{F}_{\mathrm{Hub}, K} = - \frac{\partial E_\mathrm{Hub}}{\partial \displ_K} = 
    -\sum_{i\sigma} \Bra{\psi_{i\sigma}} \frac{\partial \hat{V}_{\mathrm{Hub},\sigma}}{\partial \displ_K} \Ket{\psi_{i\sigma}} \,.
    \label{eq:Hubbard_term}
\end{equation}
In the derivation of Eq.~\eqref{eq:HF_theorem} we used Eqs.~\eqref{eq:KSeq_GS}, \eqref{eq:H_tot_GS}, and the derivative of Eq.~\eqref{eq:orthonormality}~\cite{Dalcorso:2001, Floris:2020}. Therefore, the Hubbard contribution to the force can be separated from the standard DFT force, and considered in more detail. It turns out that in practice it is more convenient to work directly with the derivative $\frac{\partial E_\mathrm{Hub}}{\partial \displ_K}$ rather than with the matrix element of $ \frac{\partial \hat{V}_{\mathrm{Hub},\sigma}}{\partial \displ_K}$~\cite{Cococcioni:2010, Cococcioni:2012, Himmetoglu:2014}, and therefore we will follow this strategy.

\subsection{Hubbard forces: The case of nonorthogonalized atomic orbitals}
\label{sec:force_atomic}

Let us consider the Hubbard force in the basis of NAO. This derivation was already presented in the case of DFT+$U$ with NC PPs~\cite{Cococcioni:2010, Cococcioni:2012}, and generalized to the case of US PPs in the context of the SIC method of Ref.~\cite{Wierzbowska:2011} which has close similarities with DFT+$U$. Here, we present a generalization to the case of DFT+$U$+$V$, in the general framework of US and PAW PPs, which is the first main result of this paper.

As was mentioned in Sec.~\ref{sec:force_general}, we need to evaluate the derivative $\frac{\partial E_\mathrm{Hub}}{\partial \displ_K}$. Using Eq.~\eqref{eq:Edftu}, we obtain:
\begin{eqnarray}
\frac{\partial E_\mathrm{Hub}}{\partial \displ_K} & = & \sum_{IJ} \sum_{\sigma m_1 m_2} \frac{\partial E_\mathrm{Hub}}{\partial n^{IJ\sigma}_{m_1 m_2}} \frac{\partial n^{IJ\sigma}_{m_1 m_2}}{\partial \displ_K} \nonumber \\
& = & \sum_I \sum_{\sigma m_1 m_2} 
U^I \left( \frac{\delta_{m_1 m_2}}{2} - n^{II \sigma}_{m_1 m_2} \right) \frac{\partial n^{II \sigma}_{m_2 m_1}}{\partial \displ_K} \nonumber \\
& & - \sum_{I} \sum_{J (J \ne I)}^* \sum_{\sigma m_1 m_2} V^{I J} 
n^{I J \sigma}_{m_1 m_2} \frac{\partial n^{J I \sigma}_{m_2 m_1}}{\partial \displ_K} \,.
\label{eq:Edftu_deriv_force}
\end{eqnarray}
Before we proceed, it is important to remark that Eq.~\eqref{eq:Edftu_deriv_force} neglects derivatives of Hubbard parameters with respect to atomic displacements, namely $\frac{\partial U^I}{\partial \displ_K} = 0$ and $\frac{\partial V^{IJ}}{\partial \displ_K} = 0$. This is a standard approximation in literature, in part due to the fact that very often empirical values of Hubbard parameters are used and hence it is not possible to evaluate such derivatives (in fact, the error made due to the ambiguity in choosing empirical $U^I$ is likely much larger than the error made due to neglecting changes in $U^I$ due to atomic displacements). However, when Hubbard parameters are computed from first principles, it is in fact possible to compute their derivatives due to atomic displacements~\cite{Kulik:2011b}. Here, for the sake of simplicity, we neglect derivatives of $U^I$ and $V^{IJ}$, but this point deserves further considerations in future studies.

Therefore, the problem is reduced to the calculation of the derivative of the generalized occupation matrix with respect to atomic displacements, which can be written as [see Eq.~\eqref{eq:occ_matrix}]:
\begin{eqnarray}
\frac{\partial n^{J I \sigma}_{m_2 m_1}}{\partial \displ_K} & = & \sum_{i} \Bigl[ 
\frac{\partial}{\partial \displ_K} \left( \bra{\psi_{i\sigma}} \hat{S} \ket{\varphi^{I}_{m_1}} \right) \bra{\varphi^{J}_{m_2}} \hat{S} \ket{\psi_{i\sigma}} \nonumber \\
& & \Bigl. + \, \bra{\psi_{i\sigma}} \hat{S} \ket{\varphi^{I}_{m_1}} \frac{\partial}{\partial \displ_K} \left( \bra{\varphi^{J}_{m_2}} \hat{S} \ket{\psi_{i\sigma}} \right) \Bigr]  \,.
\label{eq:occ_matrix_deriv_force}
\end{eqnarray}
Now the problem is to calculate the object $\frac{\partial}{\partial \displ_K} \left( \bra{\psi_{i\sigma}} \hat{S} \ket{\varphi^{I}_{m_1}} \right)$ and a similar one appearing in the equation above. 
Since KS wavefunctions do not depend explicitly on the atomic positions, the derivative $\frac{\partial \psi_{i\sigma}}{\partial \displ_K}$ is zero~\cite{Cococcioni:2010, Cococcioni:2012, Himmetoglu:2014}. Therefore, we obtain
\begin{eqnarray}
   \frac{\partial}{\partial \displ_K} \left( \bra{\psi_{i\sigma}} \hat{S} \ket{\varphi^{I}_{m_1}} \right) & = & \Bra{\psi_{i\sigma}} \frac{\partial \hat{S}}{\partial \displ_K} \Ket{\varphi^{I}_{m_1}} \nonumber \\
   & & + \Bra{\psi_{i\sigma}} \hat{S} \Ket{\frac{\partial \varphi^{I}_{m_1}}{\partial \displ_K}} \,.
   \label{eq:force_atomic_USterm}
\end{eqnarray}
The derivatives in Eq.~\eqref{eq:force_atomic_USterm} were briefly discussed in the case of the SIC method with US PPs in Ref.~\cite{Wierzbowska:2011}. Here, we use such a generalization to US (and PAW) PPs in the context of DFT+$U$+$V$ for the first time. The first term in Eq.~\eqref{eq:force_atomic_USterm} is present due to the use of US or PAW PPs and it has no counterpart in the NC PPs case~\cite{Cococcioni:2010, Cococcioni:2012, Himmetoglu:2014}.  By using Eq.~\eqref{eq:S_overlap} we obtain:
\begin{eqnarray}
   \Bra{\psi_{i\sigma}} \frac{\partial \hat{S}}{\partial \displ_K} \Ket{\varphi^{I}_{m_1}} & = &
   \sum_{L\mu\nu} q^{\gamma(L)}_{\mu\nu} \Bigl[ \Braket{\psi_{i\sigma}}{\frac{\partial \beta^L_\mu}{\partial \displ_K}} \braket{\beta^L_\nu}{ \varphi^I_{m_1}} \Bigr. \nonumber \\
   & & \hspace{0.6cm} \Bigl. + \braket{\psi_{i\sigma}}{\beta^L_\mu} \Braket{\frac{\partial \beta^L_\nu}{\partial \displ_K}}{\varphi^I_{m_1}} \Bigr] \,.
   \label{eq:force_atomic_Sderiv1}
\end{eqnarray}
Due to the locality of projector functions $\beta$ (we recall that these projector functions are different from zero only inside spheres centered on atoms) we have $\frac{\partial \beta^L_\mu}{\partial \displ_K} = \delta_{LK} \frac{\partial \beta^K_\mu}{\partial \displ_K}$, and, therefore, Eq.~\eqref{eq:force_atomic_Sderiv1} becomes:
\begin{eqnarray}
   \Bra{\psi_{i\sigma}} \frac{\partial \hat{S}}{\partial \displ_K} \Ket{\varphi^{I}_{m_1}} & = &
   \sum_{\mu\nu} q^{\gamma(K)}_{\mu\nu} \Bigl[ \Braket{\psi_{i\sigma}}{\frac{\partial \beta^K_\mu}{\partial \displ_K}} \braket{\beta^K_\nu}{ \varphi^I_{m_1}} \Bigr. \nonumber \\
   & & \hspace{0.6cm} \Bigl. + \braket{\psi_{i\sigma}}{\beta^K_\mu} \Braket{\frac{\partial \beta^K_\nu}{\partial \displ_K}}{\varphi^I_{m_1}} \Bigr] \,.
   \label{eq:force_atomic_Sderiv2}
\end{eqnarray}
It is important to stress that due to the presence of the US or PAW term given by Eq.~\eqref{eq:force_atomic_Sderiv2}, there are nonzero Hubbard forces even on non-Hubbard atoms (i.e. atoms on which we do not apply the Hubbard correction). Instead, in the case of NC PPs and NAO, Hubbard forces appear only on Hubbard atoms~\cite{Cococcioni:2010, Cococcioni:2012, Himmetoglu:2014}. 

Finally, the second term in Eq.~\eqref{eq:force_atomic_USterm} requires computing the derivative of NAO with respect to atomic displacements, $\frac{\partial \varphi^{I}_{m_1}}{\partial \displ_K}$, which is similar to the derivative $\frac{\partial \beta^K_\mu}{\partial \displ_K}$ from the implementation point of view. These objects can be efficiently computed in the reciprocal space, as was discussed in detail in Refs.~\cite{Cococcioni:2010, Cococcioni:2012, Himmetoglu:2014, Wierzbowska:2011}, and hence it will not be detailed here.

\subsection{Hubbard forces: The case of orthogonalized atomic orbitals}
\label{sec:force_orthoatomic}

In this section we present a generalization of the formalism discussed in Sec.~\ref{sec:force_atomic} to the case of OAO, which is the second main result of this paper. All the equations of Sec.~\ref{sec:force_atomic} hold also for OAO, by replacing nonorthogonalized atomic orbitals $\varphi^I_{m}(\mathbf{r})$ by orthogonalized ones $\tilde{\varphi}^I_{m}(\mathbf{r})$. The main difference with the case discussed so far is how to compute $\frac{\partial \tilde{\varphi}^{I}_{m}}{\partial \displ_K}$. In the following of this section we present the definition of OAO and then discuss how to compute their derivatives.

\subsubsection{Orthogonalized atomic orbitals}

Using the L\"owdin orthogonalization method~\cite{Lowdin:1950, Mayer:2002}, we can define OAO as:
\begin{equation}
    \tilde{\varphi}^I_{m_1}(\mathbf{r}) = \sum_{J m_2} \left(\ooverlap^{-\frac{1}{2}}\right)^{JI}_{m_2 m_1} \varphi^J_{m_2}(\mathbf{r}) \,,
    \label{eq:OAO_def}
\end{equation}
where $\ooverlap$ is the orbital overlap matrix which is defined as:
\begin{equation}
    (\ooverlap)^{IJ}_{m_1 m_2} = \bra{\varphi^I_{m_1}} \hat{S} \ket{\varphi^J_{m_2}} \,,
    \label{eq:overlap_def}
\end{equation}
where $(\ooverlap)^{IJ}_{m_1 m_2}$ is a matrix element of $\ooverlap$. Note, $(\ooverlap)^{IJ}_{m_1 m_2}$ 
can be represented as a $N \times N$ matrix by merging indices $I$ with $m_1$, and $J$ with $m_2$, where $N$ is the number of all states in the system. Therefore, in the following we will refer to $(\ooverlap)^{IJ}_{m_1 m_2}$ as a matrix. It is important to note that the following indices must be understood as being in couples, $(I,m_1)$ and $(J,m_2)$, because for different types of atoms the indices $m_1$ and $m_2$ run over different number of states. As was already anticipated in Sec.~\ref{sec:Hubbard_manifold}, here we orthogonalize \textit{all} states of \textit{all} atoms in the system. It is important to orthogonalize not only states that belong to the chosen Hubbard manifolds of each atom (e.g., $d$ or $f$ states), but also the remaining states, in order to preserve the on-site orthogonality. 

The orbital overlap matrix $\ooverlap$ is Hermitian and positive definite, therefore we can represent it as
\begin{equation}
    \ooverlap = \Umatrix \Odiag \Umatrix^\dagger \,,
    \label{eq:overlap_decomposition}
\end{equation}
where $\Odiag$ is the diagonal $N \times N$ matrix composed of eigenvalues $\{ \eigenval^I_m \}$ that are all positive, and $\Umatrix$ is the unitary $N \times N$ matrix ($\Umatrix \Umatrix^\dagger = \unit$, where $\unit$ is the unit matrix) which is formed by the corresponding eigenvectors. Using Eq.~\eqref{eq:overlap_decomposition}, we can easily compute $\ooverlap^{-\frac{1}{2}}$, which appears in Eq.~\eqref{eq:OAO_def}, as:
\begin{equation}
    \ooverlap^{-\frac{1}{2}} = \Umatrix \Odiag^{-\frac{1}{2}} \Umatrix^\dagger \,,
    \label{eq:overlap_inv}
\end{equation}
where $\Odiag^{-\frac{1}{2}}$ is the diagonal $N \times N$ matrix composed of inverse square root of eigenvalues, namely $\{ (\eigenval^I_m)^{-\frac{1}{2}} \}$.

\subsubsection{Derivatives of orthogonalized atomic orbitals}

When we replace $\varphi^I_{m}(\mathbf{r})$ by $\tilde{\varphi}^I_{m}(\mathbf{r})$ in Eq.~\eqref{eq:force_atomic_USterm}, we need to compute $\frac{\partial \tilde{\varphi}^{I}_{m_1}}{\partial \displ_K}$. According to the definition~\eqref{eq:OAO_def}, we obtain:
\begin{eqnarray}
   \frac{\partial \tilde{\varphi}^{I}_{m_1}(\mathbf{r})}{\partial \displ_K} & = &
   \sum_{J m_2} \frac{\partial}{\partial \displ_K} \Bigl[ \left(\ooverlap^{-\frac{1}{2}}\right)^{JI}_{m_2 m_1} \Bigr] \, \varphi^J_{m_2}(\mathbf{r}) \nonumber \\
   & & + \sum_{J m_2} \left(\ooverlap^{-\frac{1}{2}}\right)^{JI}_{m_2 m_1} \frac{\partial \varphi^J_{m_2}(\mathbf{r})}{\partial \displ_K} \,.
   \label{eq:OAO_deriv}
\end{eqnarray}
The second term in Eq.~\eqref{eq:OAO_deriv} is easy to compute; it is different from zero only when $J=K$ because $\varphi$ functions are atom-centered and hence $\frac{\partial \varphi^J_{m_2}(\mathbf{r})}{\partial \displ_K} = \delta_{JK} \frac{\partial \varphi^K_{m_2}(\mathbf{r})}{\partial \displ_K}$~\cite{Cococcioni:2010, Cococcioni:2012}. It is worth to note that this second term is non-zero even when $K \neq I$ but $K = J$, while in the case of NAO instead of Eq.~\eqref{eq:OAO_deriv} there is only $\frac{\partial \varphi^{I}_{m_1}(\mathbf{r})}{\partial \displ_K}$ and it equals to zero when $K \neq I$. Such a difference between OAO and NAO is important, because there are non-zero contributions to Hubbard forces for the former but not for the later when $K \neq I$. 
Now let us consider the first term in Eq.~\eqref{eq:OAO_deriv}. This term is different from zero even when $K \neq J \neq I$ due to the derivative of $\ooverlap^{-\frac{1}{2}}$ which has non-zero off-diagonal terms. Therefore, in the case of OAO, both terms in Eq.~\eqref{eq:OAO_deriv} are responsible for non-zero Hubbard forces on non-Hubbard atoms (even in the case of NC PPs) in addition to another non-zero contribution due to the use of US or PAW PPs [see discussion after Eq.~\eqref{eq:force_atomic_Sderiv2}].

The first term in Eq.~\eqref{eq:OAO_deriv} is the most challenging part and is the main focus of this paper. Indeed, at the first glance it is not trivial how to evaluate \textit{exactly} the derivative of the inverse square root of the orbital overlap matrix
\begin{equation}
    \frac{\partial}{\partial \displ_K} \Bigl[ \left(\ooverlap^{-\frac{1}{2}}\right)^{JI}_{m_2 m_1} \Bigr] \,.
    \label{eq:deriv_Oinv}
\end{equation}
In the following we discuss how this object was treated in literature so far, and we present a detailed derivation of the exact solution and its first application in the framework of DFT with extended Hubbard functionals for the evaluation of Hubbard forces (and other first-order derivatives).

\subsubsection{Derivative of $\ooverlap^{-\frac{1}{2}}$ as a solution of the Lyapunov equation}
\label{sec:Lyapunov}

In this section we discuss how to compute the exact derivative of $\ooverlap^{-\frac{1}{2}}$. This problem was already addressed in literature, and existing solutions will be briefly discussed in the following.

In Ref.~\cite{Novoselov:2015}, the authors used a representation in which the overlap matrix is close to diagonal, and hence in this case it is straightforward to evaluate the derivative in Eq.~\eqref{eq:deriv_Oinv} (see Eq.~(14) in Ref.~\cite{Novoselov:2015}). Essentially, this approximation neglects off-diagonal matrix elements by assuming that they are small, which makes the calculation of the derivatives of its powers straightforward. Such an approximation turns out to be quite good in the context of Hubbard forces using Wannier functions for systems considered in Ref.~\cite{Novoselov:2015}, however noticeable deviations were observed with respect to forces computed using finite differences. These deviations can become problematic when performing structural optimizations, where the mismatch between forces and gradients of energy can lead to instabilities of the minimization algorithms, such as e.g. the Broyden-Fletcher-Goldfarb-Shanno (BFGS) algorithm~\cite{Fletcher:1987}. Therefore, more accurate ways of computing Hubbard forces are desired. While systematic improvements of the accuracy of approximate Hubbard forces is possible (see Appendix~\ref{app:Taylor}), the exact solution is obviously desired.

In Ref.~\cite{Wu:2006}, the exact analytical formula for computing the derivative of $\ooverlap^{\frac{1}{2}}$ is presented in the context of constrained DFT to explore the diabatic potential energy curves in the Marcus theory of electron transfer (see Eq.~(11) in Ref.~\cite{Wu:2006}). Here, we are interested in the derivative of $\ooverlap^{-\frac{1}{2}}$, and hence the method of Ref.~\cite{Wu:2006} can be easily adapted to the current problem (the difference is just the sign in the power). The derivation presented in this paper was developed independently from Ref.~\cite{Wu:2006} and, at variance with what is done in that work (that only gives the final formula) will be presented in full detail~\cite{Timrov:Note:2020:Deriv_overlap_rederivation}. Authors are also aware of another independent exact derivation of the same solution~\cite{Kucukbenli:2020} as us and Ref.~\cite{Wu:2006}. This work is further motivated by the fact that it provides the first use of the analytical expression of the derivative of $\ooverlap^{-\frac{1}{2}}$ in the context of Hubbard-corrected DFT.

In the following, we discuss in detail the derivation of the exact analytical formula for evaluating Eq.~\eqref{eq:deriv_Oinv}. Taking the derivative of both members of the identity $\InvSqrtO \InvSqrtO = \ooverlap^{-1}$, it is easy to find:
\begin{equation}
    \InvSqrtO \frac{\partial \InvSqrtO}{\partial \displ_K} + \frac{\partial \InvSqrtO}{\partial \displ_K} \InvSqrtO = \DinvO_K \,,
    \label{eq:Lyapunov_eq}
\end{equation}
where
\begin{equation}
    \DinvO_K \equiv \frac{\partial \ooverlap^{-1}}{\partial \displ_K} = - \ooverlap^{-1} \frac{\partial \ooverlap}{\partial \displ_K} \ooverlap^{-1} \,.
    \label{eq:W_def}
\end{equation}
Equation~\eqref{eq:W_def} was obtained by taking a derivative of the identity $\ooverlap \ooverlap^{-1} = \unit$, and then by multiplying the resulting equation by $\ooverlap^{-1}$ on the left-hand side.

Equation~\eqref{eq:Lyapunov_eq} can be identified as a type of \textit{Lyapunov equation}~\cite{Lyapunov:1948}, or as a particular case of the \textit{Sylvester equation}~\cite{Bartels:1972}. Numerical solutions to Eq.~\eqref{eq:Lyapunov_eq} can be obtained with the help of the various algorithms as variants of the classical Bartels–Stewart algorithm~\cite{Bartels:1972} requiring a QR factorization of the matrix $\InvSqrtO$. Importantly, these algorithms do not require the diagonalization of $\InvSqrtO$ which can be computationally very expensive for matrices of large size. Unfortunately, all implementations available at the moment as part of standard linear algebra libraries are very limited in terms of size of the matrix $\InvSqrtO$, and therefore of little (if any) use in applications to electronic structure problems.
However, for all cases in which one can assume that the diagonalization of the matrices does not introduce a major computational bottleneck, an exact solution to Eq.~\eqref{eq:Lyapunov_eq} can be obtained through a closed-form expression~\cite{Wu:2006}. As shown in Appendix \ref{app:Lyap_solution}, under conditions fulfilled by the eigenvalues of the matrix $\ooverlap$, we can write the formal solution to Eq.~\eqref{eq:Lyapunov_eq} as follows: 
\begin{equation}
   \frac{\partial \InvSqrtO}{\partial \displ_K} = \int\limits_0^\infty e^{- \ttt \, \InvSqrtO} \, \DinvO_K \, e^{- \ttt \, \InvSqrtO} \, d\ttt \,,  
   \label{eq:dS1}
\end{equation}
where $\ttt$ is an auxiliary (scalar) integration variable. Using Eq.~\eqref{eq:overlap_inv}, we can rewrite Eq.~\eqref{eq:dS1} as:
\begin{equation}
   \frac{\partial \InvSqrtO}{\partial \displ_K} = \Umatrix \left( \int\limits_0^\infty e^{- \ttt \, \Odiag^{-\frac{1}{2}}} \, \tilde{\DinvO}_K \, e^{- \ttt \, \Odiag^{-\frac{1}{2}}} \, d\ttt \right) \Umatrix^\dagger \,,  
   \label{eq:dS2} 
\end{equation}
where
\begin{equation}
    \tilde{\DinvO}_K = \Umatrix^\dagger \DinvO_K \Umatrix \,.
    \label{eq:Wtilde}
\end{equation}
We can write even more useful result by actually computing the integral in Eq.~\eqref{eq:dS2}. The matrix element of the term in brackets of Eq.~\eqref{eq:dS2} reads:
\begin{eqnarray}
    & & \left( \int\limits_0^\infty e^{- \ttt \, \Odiag^{-\frac{1}{2}}} \, \tilde{\DinvO}_K \, e^{- \ttt \, \Odiag^{-\frac{1}{2}}} \, d\ttt \right)^{IJ}_{m_1 m_2} \nonumber \\
    & & \hspace{0.7cm} = \int\limits_0^\infty e^{- \ttt \, (\eigenval^I_{m_1})^{-\frac{1}{2}}} \, (\tilde{\DinvO}_K)^{IJ}_{m_1 m_2} \, e^{- \ttt \, (\eigenval^J_{m_2})^{-\frac{1}{2}}} \, d\ttt \nonumber \\
    & & \hspace{0.7cm} = \frac{(\tilde{\DinvO}_K)^{IJ}_{m_1 m_2}}{(\eigenval^I_{m_1})^{-\frac{1}{2}} + (\eigenval^J_{m_2})^{-\frac{1}{2}}} \,.
    \label{eq:dS3}
\end{eqnarray}
We note that a solution given by Eqs.~\eqref{eq:dS2}--\eqref{eq:dS3} corresponds to a similar result in Ref.~\cite{Wu:2006}. Using Eqs.~\eqref{eq:dS2} and \eqref{eq:dS3}, we obtain:
\begin{eqnarray}
    & & \left(\frac{\partial \InvSqrtO}{\partial \displ_K}\right)^{IJ}_{m_1 m_2}  \nonumber \\
    & & = \sum_{L m_3, M m_4} \frac{ \bigl(\Umatrix\bigr)^{IL}_{m_1 m_3} (\tilde{\DinvO}_K)^{LM}_{m_3 m_4} \bigl(\Umatrix^\dagger \bigr)^{MJ}_{m_4 m_2}}{(\eigenval^L_{m_3})^{-\frac{1}{2}} + (\eigenval^M_{m_4})^{-\frac{1}{2}}} \,.
    \label{eq:dS4}
\end{eqnarray}
Finally, using Eqs.~\eqref{eq:W_def} and \eqref{eq:Wtilde}, and the fact that $\ooverlap^{-1} = \Umatrix \Odiag^{-1} \Umatrix^\dagger$, Eq.~\eqref{eq:dS4} can be rewritten in the compact form:
\begin{eqnarray}
   & & \left(\frac{\partial \InvSqrtO}{\partial \displ_K}\right)^{IJ}_{m_1 m_2} \nonumber \\
   & & = - \sum_{L m_3, M m_4} \frac{\bigl(\Umatrix\bigr)^{IL}_{m_1 m_3} \left(\Umatrix^\dagger \frac{\partial \ooverlap}{\partial \displ_K} \Umatrix\right)^{LM}_{m_3 m_4} \bigl(\Umatrix^\dagger \bigr)^{MJ}_{m_4 m_2}}{\eigenval^L_{m_3} (\eigenval^M_{m_4})^{\frac{1}{2}} + \eigenval^M_{m_4} (\eigenval^L_{m_3})^{\frac{1}{2}}} \,.
   \label{eq:dS5}
\end{eqnarray}
Equation~\eqref{eq:dS5} is exact, and it constitutes a closed form of the solution of the Lyapunov equation~\eqref{eq:Lyapunov_eq}. It can be seen from Eq.~\eqref{eq:dS5} that the only ingredients that are needed to compute the derivative of $\ooverlap^{-\frac{1}{2}}$ are the eigenvalues and eigenvectors of the overlap matrix $\ooverlap$, and the derivative of this matrix, $\frac{\partial \ooverlap}{\partial \displ_K}$. The latter can be easily computed using the definition of the overlap matrix~\eqref{eq:overlap_def}:
\begin{eqnarray}
    \frac{\partial \ooverlap}{\partial \displ_K} & = & 
    \Bra{\frac{\partial \varphi^I_{m_1}}{\partial \displ_K}} \hat{S} \Ket{\varphi^J_{m_2}} + 
    \Bra{\varphi^I_{m_1}} \hat{S} \Ket{\frac{\partial \varphi^J_{m_2}}{\partial \displ_K}} \nonumber \\
    & & + \, \Bra{\varphi^I_{m_1}} \frac{\partial \hat{S}}{\partial \displ_K} \Ket{\varphi^J_{m_2}} \,.
    \label{eq:overlap_deriv}
\end{eqnarray}
The first two terms in Eq.~\eqref{eq:overlap_deriv} are computed using the localized character of atomic orbitals:
\begin{eqnarray}
   & & \Bra{\frac{\partial \varphi^I_{m_1}}{\partial \displ_K}} \hat{S} \Ket{\varphi^J_{m_2}} + 
       \Bra{\varphi^I_{m_1}} \hat{S} \Ket{\frac{\partial \varphi^J_{m_2}}{\partial \displ_K}} \nonumber \\
   & & = \delta_{IK} \Bra{\frac{\partial \varphi^K_{m_1}}{\partial \displ_K}} \hat{S} \Ket{\varphi^J_{m_2}} + \delta_{JK} \Bra{\varphi^I_{m_1}} \hat{S} \Ket{\frac{\partial \varphi^K_{m_2}}{\partial \displ_K}} \,,
\end{eqnarray}
and the last term in Eq.~\eqref{eq:overlap_deriv} is computed in a way similar to Eq.~\eqref{eq:force_atomic_Sderiv2}.

It is worth to point out that the method presented here for the calculation of the first derivative of $\ooverlap^{-\frac{1}{2}}$ can be generalized to higher-order derivatives. In particular, the second derivative of $\ooverlap^{-\frac{1}{2}}$ can be computed by differentiating Eq.~\eqref{eq:Lyapunov_eq} and by writing a solution of the resulting equation in a closed form similarly to Eq.~\eqref{eq:dS1}. As an example, such a computation will be useful for the generalization of the density-functional perturbation theory with the Hubbard $U$ correction (the so-called DFPT+$U$ approach) for calculation of phonons, which requires second-order derivatives of the occupation matrix for the calculation of the matrix of interatomic force constants~\cite{Floris:2020}.

Finally, the computational scaling of Hubbard forces using OAO is compared to that using NAO and is presented in Appendix~\ref{sec:Scaling}.

\section{Technical details}
\label{sec:technical_details}

The formalism for computing Hubbard forces presented in Sec.~\ref{sec:Theory} has been implemented in the \textsc{Quantum ESPRESSO} distribution~\cite{Giannozzi:2009, Giannozzi:2017, Giannozzi:2020}, and it is publicly available to the community~\cite{QuantumESPRESSO:website}. In order to benchmark this implementation, we consider the case of NiO. In this section we review the technical settings of these calculations.

We have used the experimental lattice parameter $a = 4.17$~\AA \, for the rock-salt crystal structure of antiferromagnetic NiO~\cite{Bartel:1971}. We have used the GGA for the exchange-correlation functional constructed with the PBEsol prescription~\cite{Perdew:2008}. Pseudopotentials were taken from the SSSP library~1.1 efficiency~\cite{Prandini:2018, MaterialsCloud}: for Ni we have used the US PP from the GBRV library~1.4~\cite{Garrity:2014} (\texttt{ni\_pbesol\_v1.4.uspp.F.UPF}), and for O we have used the PAW PP from the Pslibrary library~0.3.1~\cite{Kucukbenli:2014} (\texttt{O.pbesol-n-kjpaw\_psl.0.1.UPF}). KS wavefunctions and potentials were expanded in plane waves using the kinetic-energy cutoffs of 80 and 640~Ry, respectively. The Brillouin zone has been sampled with a uniform $8 \times 8 \times 8$ $\mathbf{k}$ point mesh centered at the $\Gamma$~point. The accuracy on the computed forces is better than $10^{-5}$~Ry/bohr.

The Hubbard parameters were computed using the \texttt{HP} code which is based on DFPT~\cite{Timrov:2018, Timrov:2020b}. We have used the $\mathbf{k}$ and $\mathbf{q}$ point meshes of size $8 \times 8 \times 8$ and $5 \times 5 \times 5$, respectively, which give an accuracy of 0.01~eV for the computed values of $U$ and $V$. Hubbard parameters were computed using the self-consistent procedure, which is described in detail in Refs.~\cite{Hsu:2009, Cococcioni:2019, Timrov:2020b}. In the framework of DFT+$U$+$V$, we obtained the following values: using OAO, $U_\mathrm{OAO} = 7.43$~eV for Ni($3d$) states, and $V_\mathrm{OAO} = 0.37$~eV between Ni($3d$) and O($2p$) states; using NAO, $U_\mathrm{NAO} = 6.76$~eV and $V_\mathrm{NAO} = 1.25$~eV.

For the benchmarking purposes, we evaluated the total force using numerical differentiation (finite differences) of the total energy. We used the symmetric difference quotient, with the atomic displacement parameter of $5 \times 10^{-3}$~bohr along each Cartesian direction. 

The data used to produce the results of this work are available at the Materials Cloud Archive~\cite{MaterialsCloudArchive2020}.

\section{Results}
\label{sec:results}

We now proceed to validate the correctness of implementation of Hubbard forces with OAO by considering NiO as a test case, as it is one of the most studied materials using the DFT + Hubbard scheme, and several theoretical and experimental studies are available in literature. Here we focus on the DFT+$U$+$V$ case, which is the most general one; similar trends and results were obtained also for the DFT+$U$ case, but they will not be discussed in the following.

Below a Ne\'el temperature of 523~K, NiO has an antiferromagnetic ordering of type~II (AFII), where ferromagnetic (111) Ni planes alternate with opposite magnetization along the [111] direction~\cite{Campo:2010}. This magnetic ordering is compatible with rhombohedral symmetry, and hence the crystal structure can be modelled using a primitive cell with four atoms, which we label as Ni$_1$, Ni$_2$, O$_1$, and O$_2$. To impose the AFII magnetic ordering, we assign to Ni$_1$ a spin up polarization and for Ni$_2$ a spin down one, while O$_1$ and O$_2$ have zero net spin polarization. The atomic positions are as follows: Ni$_1$ is at $(0, 0, 0)$, Ni$_2$ is at $(1/2, 1/2, 0)$, O$_1$ is at $(1/2, 0, 0)$, and O$_2$ is at $(1/2, 1/2, 1/2)$ in the Cartesian framework in units of the lattice parameter $a$ (see Ref.~\cite{Cococcioni:2005} for more details). Since all atoms sit in high-symmetry positions, at equilibrium the forces are zero on all atoms and in all directions. This is an ideal test case for the current study, because now we can displace atoms and compute non-zero forces acting on them, and thus we can benchmark the accuracy of our implementation of total and Hubbard forces with OAO.

Before we proceed with the calculation of forces, for the sake of completeness we summarize here basic quantities obtained in our study with DFT+$U$+$V$ and OAO. Using the self-consistent Hubbard parameters reported in Sec.~\ref{sec:technical_details}, we obtain a band gap of 3.14~eV, and the magnetic moments of Ni of 1.72~$\mu_\mathrm{B}$ -- these values are in good agreement with previous theoretical and experimental studies~\cite{Campo:2010}. We will not elaborate more on this, since this is not the focus of the current study.

\subsection{Benchmark of the total force using OAO}

In this section we validate the implementation of the Hubbard forces [Eq.~\eqref{eq:Hubbard_term}] using OAO by comparing the total force [Eq.~\eqref{eq:HF_theorem}] computed using our analytical formulas and by using finite differences of the total energy. Our benchmark procedure is along the same lines as in Refs.~\cite{Novoselov:2015, Tran:2008, Wierzbowska:2011}.

\begin{figure}[t]
\begin{center}
 \includegraphics[width=0.97\linewidth]{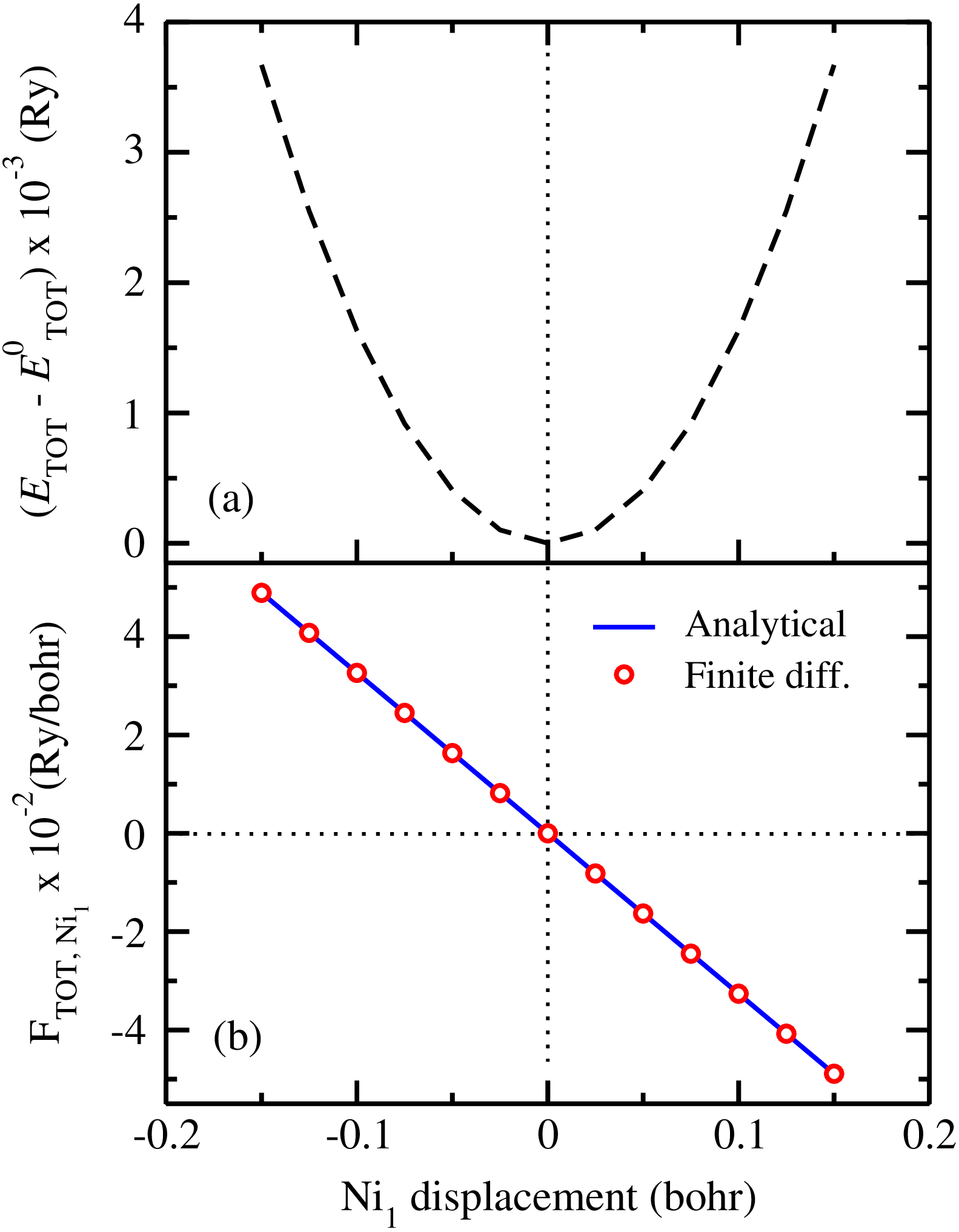}
 \caption{(a)~Total energy $E_\mathrm{TOT}$ [see Eq.~\eqref{eq:Edft_plus_u}] as a function of the displacement on Ni$_1$ atom along the [111] direction. $E^0_\mathrm{TOT}$ is the total energy computed at zero displacement of the Ni$_1$ atom. (b)~Total force $\mathrm{F}_{\mathrm{TOT}, \mathrm{Ni}_1}$ acting on the Ni$_1$ atom when it is displaced along the [111] direction from its high-symmetry position (0,0,0). Blue line represents the total force computed using the analytical expression given by Eq.~\eqref{eq:HF_theorem}, and red empty circles represent the total force computed using finite differences of the total energy shown in panel~(a). Data on panels~(a) and (b) were obtained using OAO, and using Hubbard parameters $U_\mathrm{OAO}$ and $V_\mathrm{OAO}$.}
\label{fig1}
\end{center}
\end{figure}

Figure~\ref{fig1} shows the total energy and total force acting on the Ni$_1$ atom as a function of the magnitude of its displacement along the [111] direction from its equilibrium position (0,0,0), obtained using OAO. The components of the force are equal along three Cartesian directions due to symmetry. The total force is computed as a square root of squared Cartesian components, and we also preserve the sign of the total force in order to highlight its direction. The ``analytical'' total force (blue solid line) was computed using Eq.~\eqref{eq:HF_theorem}, i.e. the standard DFT force plus the Hubbard force using OAO as defined in Eq.~\eqref{eq:Hubbard_term}, which is based on the exact calculation of various derivatives and in particular that of the $\ooverlap^{-\frac{1}{2}}$ matrix as discussed in Sec.~\ref{sec:Lyapunov}. We also computed the total force using numerical differentiation (finite differences) of the total energy [see Fig.~\ref{fig1}~(a)] and the result is shown in Fig.~\ref{fig1}~(b) (red empty circles). It is easy to see from Fig.~\ref{fig1}~(b) that the analytical and numerical differentiation give exactly the same total force for various magnitudes of the displacement of Ni$_1$ atom, which validates the correctness of formulas and of their implementation.

From Fig.~\ref{fig1}~(b) it is difficult to extract information about the precision of agreement between the analytical and numerical solutions (we just visually see that the results are on top of each other). In fact, in Ref.~\cite{Novoselov:2015} the authors also considered NiO as a test case to benchmark the implementation of Hubbard forces, and they used an approximate expression for the derivative of $\ooverlap^{-\frac{1}{2}}$ in the framework of Wannier functions; although the results were very good, the agreement between analytical and numerical forces was not perfect. Here, in order to elaborate more on the accuracy of our formalism, we present a comparison of the analytical and numerical results for the total force for the largest displacement considered here (+0.15~bohr). From Table~\ref{tab1} we can see that the total forces agree with the accuracy of $10^{-6}$~Ry/bohr, which is remarkable and further validates the correctness and high accuracy of our implementation of analytical formulas for the total and Hubbard forces. As a side note, we see from Table~\ref{tab1} that the total force acting on Ni$_1$ is negative, while total forces acting on all other atoms are all positive; this means that all atoms (except Ni$_1$) counteract to the displaced Ni$_1$ atom, and the sum of all forces is zero.

\begin{table}[t]
 \begin{center}
  \begin{tabular}{ccc}
    \hline\hline
     \parbox{2.5cm}{Atom}   &    \parbox{2.5cm}{F$_\mathrm{finite\,\,diff.}^\mathrm{OAO}$}  &  \parbox{2.5cm}{F$_\mathrm{analytical}^\mathrm{OAO}$}   \\ \hline
     Ni$_1$       &    $-0.048891$   &    $-0.048891$     \\
     Ni$_2$       &    $0.026303$    &    $0.026303$      \\
     O$_1$        &    $0.012951$    &    $0.012950$      \\
     O$_2$        &    $0.009639$    &    $0.009638$      \\
     \hline\hline
  \end{tabular}    
 \end{center}
\caption{Comparison of the total force (in Ry/bohr) acting on atoms upon a displacement of the Ni$_1$ atom by $+0.15$~bohr along the [111] direction. The results are presented for the case of finite differences of the total energy and analytical implementation of forces using OAO, and using Hubbard parameters $U_\mathrm{OAO}$ and $V_\mathrm{OAO}$.}
\label{tab1}
\end{table}

\subsection{Hubbard contribution to the total force and the importance of the derivative of $\ooverlap^{-\frac{1}{2}}$}

In this section we discuss the contribution of the Hubbard force to the total force computed using analytical expressions given by Eqs.~\eqref{eq:HF_theorem} and \eqref{eq:Hubbard_term}, using OAO as a localized basis set. Moreover, we investigate the importance of taking into account the derivative of $\ooverlap^{-\frac{1}{2}}$ in Eq.~\eqref{eq:OAO_deriv}.

\begin{table*}[t]
 \begin{center}
  \begin{tabular}{cccccc}
    \hline\hline
     \multirow{2}{*}{\parbox{2cm}{Atom}}   &  \multicolumn{2}{c}{\parbox{4.5cm}{F$_\mathrm{analytical}^\mathrm{OAO}$ (exact)}}  & & \multicolumn{2}{c}{\parbox{4.5cm}{F$_\mathrm{analytical}^\mathrm{OAO}$ (approximate)}}  \\ \cline{2-3} \cline{5-6}
                  &   Total force      &    Hubbard force    & &   Total force      &    Hubbard force     \\ \hline
     Ni$_1$       &   $-0.048891$      &     $0.001652$      & &   $-0.050309$      &    $0.000234$        \\
     Ni$_2$       &   $0.026303$       &    $-0.010003$      & &    $0.026656$      &   $-0.009650$        \\
     O$_1$        &   $0.012950$       &     $0.000204$      & &    $0.013427$      &    $0.000680$        \\
     O$_2$        &   $0.009638$       &     $0.000050$      & &    $0.010226$      &    $0.000638$        \\
     \hline\hline
  \end{tabular}    
 \end{center}
\caption{Total and Hubbard forces (in Ry/bohr) computed using analytical formulas Eqs.~\eqref{eq:HF_theorem} and \eqref{eq:Hubbard_term} for the Ni$_1$ atom displacement of $+0.15$~bohr along the [111] direction. Second and third columns correspond to exact forces computed using Eq.~\eqref{eq:OAO_deriv}, while the fours and fifth columns correspond to approximate forces computed by neglecting the first term in Eq.~\eqref{eq:OAO_deriv}. In all cases OAO were used for the calculation of forces together with the Hubbard parameters $U_\mathrm{OAO}$ and $V_\mathrm{OAO}$.}
\label{tab2}
\end{table*}

Table~\ref{tab2} presents forces computed using the exact (full) expression for the derivative of OAO given by Eq.~\eqref{eq:OAO_deriv}, and using an approximate expression which neglects entirely the derivative of $\ooverlap^{-\frac{1}{2}}$ in Eq.~\eqref{eq:OAO_deriv} (i.e. the first term). In both cases Hubbard parameters $U_\mathrm{OAO}$ and $V_\mathrm{OAO}$ were used. We discuss first the exact forces and then the approximate ones.

As can be seen from Table~\ref{tab2} (second and third columns), there are non-zero Hubbard forces acting on all atoms when we displace Ni$_1$. As was already discussed in Sec.~\ref{sec:Theory}, the non-zero Hubbard forces on non-Hubbard atoms O$_1$ and O$_2$ is a consequence of two factors: (i)~the use of US and PAW PPs [see Eq.~\eqref{eq:force_atomic_Sderiv2}], and (ii)~the use of OAO [see Eq.~\eqref{eq:OAO_deriv} and the discussion after this equation]. Interestingly, the Hubbard force acting on Ni$_2$ is one order of magnitude larger than the Hubbard force acting on the displaced atom Ni$_1$. Moreover, the signs of Hubbard forces acting on different atoms are different and it is not trivial to guess them. These two latter observations are a consequence of a complex interplay between different contributions entering in the definition of the Hubbard force [see in particular Eqs.~\eqref{eq:force_atomic_Sderiv2} and \eqref{eq:OAO_deriv}]. Now let us compare the magnitude of the Hubbard forces with respect to the total forces. For Ni$_1$ the Hubbard force constitutes 3.4\% of the total force, for Ni$_2$ it is 38.0\%, for O$_1$ it is 1.6\%, and for O$_2$ it is 0.5\%. Therefore, while Hubbard forces are quite negligible for some atoms (e.g. O$_2$), for other atoms the Hubbard contribution to the total force is very large (e.g. Ni$_2$), and hence Hubbard forces must always be computed and added to the total force.

In this work, large effort was dedicated to the derivation of the exact expression for the derivative of $\ooverlap^{-\frac{1}{2}}$, and therefore it would be interesting and instructive to investigate the importance of this contribution in Eq.~\eqref{eq:OAO_deriv}, and thus its significance for the Hubbard and total forces. In Table~\ref{tab2} (fourth and fifth columns) we show the Hubbard and total forces acting on atoms computed by entirely neglecting the derivative of $\ooverlap^{-\frac{1}{2}}$. We can see that the total force is changed with respect to the exact total force (second column in Table~\ref{tab2}) by 2.9\% for Ni$_1$, 1.3\% for Ni$_2$, 3.7\% for O$_1$, and 6.1\% for O$_2$. These are rather significant deviations. Moreover, such errors of several percentages on total forces may lead to instabilities of structural optimization algorithms (such as BFGS~\cite{Fletcher:1987}) which are based on the comparison of the total forces and gradients of total energy. Hence, the contribution from the derivative of $\ooverlap^{-\frac{1}{2}}$ is relevant and must be always included. The Hubbard forces are affected markedly if we neglect the contribution from the derivative of $\ooverlap^{-\frac{1}{2}}$: The largest deviation is for O$_2$ and there is a factor of $\sim 12$ difference with respect to the exact Hubbard force, while the smallest change is for Ni$_2$ and it is 3.5\%. This finding underlines once more time the importance of including the derivative of $\ooverlap^{-\frac{1}{2}}$ in the formalism. For some applications (e.g. linear scaling algorithms) it might be of interest to use some approximate ways in computing the derivative of $\ooverlap^{-\frac{1}{2}}$ (such as the one described in Ref.~\cite{Novoselov:2015} or in Appendix~\ref{app:Taylor}), but the application of the current formulation for the exact evaluation of $\ooverlap^{-\frac{1}{2}}$ as a solution of the Lyapunov equation is feasible and convenient for quite large systems (from several tens to a few hundreds of atoms).

\subsection{OAO versus NAO forces}

In this section we present a comparison of the total and Hubbard forces computed using OAO and NAO. 
As was mentioned in the introduction (Sec.~\ref{sec:Introduction}), it is important to keep the consistency of the Hubbard manifolds which are used for the calculation of Hubbard parameters, total energy, forces, and various other properties. Table~\ref{tab3} presents a comparison for the fully consistent cases of OAO and NAO (i.e. when forces and Hubbard parameters are computed by using consistently the same Hubbard manifold), and for the mixed case when NAO are used to compute forces while Hubbard parameters were obtained using OAO. We stress that this latter case is presented only for demonstrative purposes, while in practice full consistency of Hubbard manifolds should be used. 
\begin{table*}[t]
 \begin{center}
  \begin{tabular}{ccccccccc}
    \hline\hline
     \multirow{2}{*}{\parbox{2cm}{Atom}}   &  \multicolumn{2}{c}{\parbox{4.5cm}{F$_\mathrm{analytical}^\mathrm{OAO}$ ($U_\mathrm{OAO} \, \& \, V_\mathrm{OAO}$)}}  & & \multicolumn{2}{c}{\parbox{4.5cm}{F$_\mathrm{analytical}^\mathrm{NAO}$ ($U_\mathrm{NAO} \, \& \, V_\mathrm{NAO}$)}} & & \multicolumn{2}{c}{\parbox{4.5cm}{F$_\mathrm{analytical}^\mathrm{NAO}$ ($U_\mathrm{OAO} \, \& \, V_\mathrm{OAO}$)}}  \\ \cline{2-3} \cline{5-6} \cline{8-9}
                  &   Total force      &    Hubbard force    & &   Total force      &    Hubbard force  & &   Total force      &    Hubbard force   \\ \hline
     Ni$_1$       &   $-0.048891$      &     $0.001652$      & &   $-0.043358$      &    $0.004127$     & &   $-0.046051$      &    $0.004006$      \\
     Ni$_2$       &   $0.026303$       &    $-0.010003$      & &    $0.023360$      &   $-0.009405$     & &    $0.024727$      &   $-0.010772$       \\
     O$_1$        &   $0.012950$       &     $0.000204$      & &    $0.011567$      &   $-0.000117$     & &    $0.012207$      &   $-0.000056$       \\
     O$_2$        &   $0.009638$       &     $0.000050$      & &    $0.008431$      &   $-0.000101$     & &    $0.009117$      &   $-0.000032$       \\
     \hline\hline
  \end{tabular}    
 \end{center}
\caption{Total and Hubbard forces (in Ry/bohr) computed using analytical formulas Eqs.~\eqref{eq:HF_theorem} and \eqref{eq:Hubbard_term} for the Ni$_1$ atom displacement of $+0.15$~bohr along the [111] direction. Second and third columns correspond to forces computed using OAO with $U_\mathrm{OAO}$ and $V_\mathrm{OAO}$; fourth and fifth columns correspond to forces computed using NAO with $U_\mathrm{NAO}$ and $V_\mathrm{NAO}$; and, sixth and seventh columns correspond to forces computed using NAO with $U_\mathrm{OAO}$ and $V_\mathrm{OAO}$.}
\label{tab3}
\end{table*}

As can be seen from Table~\ref{tab3} the differences between fully consistent total forces obtained using OAO and NAO (second and fourth columns) vary in the range from 11\% to 14\%. This is a rather significant difference, which might be very relevant when optimizing atomic positions for complex transition-metal oxides. As expected, the Hubbard forces in these two cases are very different (see columns three and five in Table~\ref{tab3}): Not only the absolute values of Hubbard forces differ in OAO and NAO, but even the sign of Hubbard forces for some atoms (O$_1$ and O$_2$ in this case) are opposite to each other~\cite{Timrov:Note:2020:forces}. This latter finding is obviously related to the extra contributions coming from the $\ooverlap^{-\frac{1}{2}}$ matrix in Eq.~\eqref{eq:OAO_deriv} when using OAO.

The origin of the difference between fully consistent total forces is twofold: $(i)$~different Hubbard parameters are used in two calculations, $(ii)$~different Hubbard manifolds are used when computing total energies and forces. In order to disentangle these two effects, we performed an additional calculation: total energies and forces are computed using NAO, while Hubbard parameters are obtained using OAO (see last two columns in Table~\ref{tab3}). We can see that in this case the difference between OAO and NAO total forces (second and sixth columns) is reduced down to 5--6\%. Therefore, such a residual difference is due to the use of different Hubbard manifolds for computing total energies and forces. Furthermore, we can see that in this mixed case also the Hubbard forces are changed in such a way that they become closer to the fully consistent OAO Hubbard forces (compare columns three, five, and seven in Table~\ref{tab3}).

Therefore, such a comparative analysis highlights that it is important to keep the full consistency when computing Hubbard parameters, total energies, forces, and other properties of materials. Most importantly, such a consistency must be used from the conceptual point of view: Mixing different Hubbard manifolds is not justified, and ultimately it can lead to unpredictable behavior of DFT+$U$ and DFT+$U$+$V$ simulations.

Finally, it would be very useful and important to compare various materials' properties using NAO and OAO, and in particular compare atomic positions after structural optimization with available x-ray diffraction data. This task requires systematic study for various types of materials which have different types of interactions (ionic, covalent, or mixed) to see trends, but this is not the objective of this paper and it is beyond its scope. However, we want to stress that such a comparison of NAO and OAO is possible now thanks to the current work, and in fact there are already ongoing efforts the results of which will be presented elsewhere.

\section{Conclusions}
\label{sec:conclusions}

We have presented a detailed derivation for the exact expression of the contribution to Pulay  forces in the framework of extended Hubbard functionals (DFT+$U$+$V$~\cite{Campo:2010}) that originate from the use of orthogonalized atomic orbitals. At variance with the use of nonorthogonalized orbitals, a major difficulty arises in this case due to the need to compute the derivative of the inverse square root of the orbital overlap matrix $\ooverlap^{-\frac{1}{2}}$. Similarly to Ref.~\cite{Wu:2006}, we have developed and presented in full detail the derivation of a closed form expression for the derivative of $\ooverlap^{-\frac{1}{2}}$ via an integral representation of the solution of the associated Lyapunov equation, where, for the sake of generality, all equations are written for the case of ultrasoft pseudopotentials and projector-augmented-wave method.

The implementation of Hubbard forces using orthogonalized atomic orbitals is benchmarked versus finite differences of total energies for NiO, and excellent agreement between forces is obtained, which validates the correctness of the analytical formulas and of the implementation. In addition, we show that the contribution from the derivative of $\ooverlap^{-\frac{1}{2}}$ to the total force is quite significant (from 1\% to 6\% in the case of NiO), and hence should always be included for the sake of consistency and high accuracy. 

Furthermore, we have compared total and Hubbard forces computed using orthogonalized and nonorthogonalized atomic orbitals using Hubbard parameters computed with the respective set of orbitals (i.e. using respective Hubbard manifolds). We have found that the differences in total forces is as large as 11--14\% in NiO, which is in part related to the use of different Hubbard parameters and in part due to the use of different Hubbard manifolds when computing Hubbard and total forces. This finding highlights the importance of maintaining the consistency of the Hubbard manifold and of the Hubbard parameters when computing various materials' properties such as total energy, forces, stress, and phonons, to name a few.

Finally, the current formalism for computing Hubbard forces using orthogonalized atomic orbitals has been implemented in the open-source \textsc{Quantum ESPRESSO} distribution~\cite{Giannozzi:2009, Giannozzi:2017, Giannozzi:2020} and is freely available to the community at large. The computational scaling of this formalism is higher than that when using nonorthogonalized atomic orbitals, however through the use of standardized mathematical libraries (BLAS and LAPACK~\cite{Anderson:1999}) and effective parallelization strategies (across plane waves, $\mathbf{k}$ points, and bands) large speedups are achieved that make the overhead of no concern. Last, we believe that this work opens up avenues for very accurate geometry optimizations for transition-metal and rare-earth compounds by taking into account complex hybridization effects between neighboring sites via the orbital overlap matrix and subsequently using it in the framework of DFT+$U$+$V$ which has proven to be very effective in complex materials~\cite{Kulik:2011,Cococcioni:2019, Ricca:2020}.

\section{ACKNOWLEDGMENTS}

M.C. is grateful to Emine K\"{u}\c{c}\"{u}kbenli and Stefano de Gironcoli for fruitful discussions at an early stage of the project. This research was supported by the Swiss National Science Foundation (SNSF), through Grant No.~200021-179138, and its National Centre of Competence in Research (NCCR) MARVEL. F.A. acknowledges financial support from the European H2020 Intersect project through Grant No.~814487 and the NFFA-Europe Transnational Access Activity through Grant No.~654360. M.C. acknowledges partial support from the EU-H2020 research and innovation program under Grant Agreement No.~654360 within the framework of the NFFA-Europe Transnational Access Activity. Computer time was provided by the Swiss National Supercomputing Centre (CSCS) under Project No.~s836.

\appendix

\section{Derivative of $\ooverlap^{-\frac{1}{2}}$ via Taylor series}
\label{app:Taylor}

In this Appendix we present an alternative method for the calculation of the derivative of $\ooverlap^{-\frac{1}{2}}$, which is based on the expansion of $\ooverlap$ in Taylor series and then computing its derivative. At variance with the exact solution presented in Sec.~\ref{sec:Lyapunov}, the use of a Taylor series is by its nature approximate and, thus, it requires convergence checks with respect to the number of terms to be included in the series. However, this approach is still attractive in cases where the matrix $\ooverlap$ is very large, as means to avoid the otherwise expensive diagonalization. This is a typical situation for linear-scaling quantum chemistry algorithms~\cite{TheDaltonCode}, as well as DFT codes that exploit extensively sparse linear algebra for calculations on large systems~\cite{ONETEP}. In this case, the use of localized basis functions leads to diagonally dominant overlap matrices, and therefore the Taylor series can be designed with optimal convergence properties, as explained in what follows.
We start by writing the overlap matrix [defined in Eq.~\eqref{eq:overlap_def}] as:
\begin{equation}
    \ooverlap = \mu \left( \frac{\ooverlap}{\mu} \right) = \mu \left( \unit + \Amatrix \right) \,,
    \label{eq:Otaylor1}
\end{equation}
where
\begin{equation}
    \Amatrix = \frac{\ooverlap}{\mu} - \unit \,.
    \label{eq:Taylor_Amatrix}
\end{equation}
Here, $\mu = \mathrm{max}[\eigenval^I_m]$ is the maximum eigenvalue of $\ooverlap$ [see Eq.~\eqref{eq:overlap_decomposition}], $\unit$ is the unit matrix, and $\Amatrix$ is an auxiliary matrix with a spectral radius (i.e., the magnitude of its largest eigenvalue) strictly smaller than 1. The representation of $\ooverlap$ given by Eqs.~\eqref{eq:Otaylor1} and \eqref{eq:Taylor_Amatrix} is particularly useful for situations when the orbitals are not normalized to 1.

The Taylor series for $\ooverlap^{-\frac{1}{2}}$ in terms of $\Amatrix$ can then be obtained using Eq.~\eqref{eq:Otaylor1}, namely 
\begin{eqnarray}
    \ooverlap^{-\frac{1}{2}} & = & \mu^{-\frac{1}{2}} \left( \unit + \Amatrix \right)^{-\frac{1}{2}} \nonumber \\
    & = & \mu^{-\frac{1}{2}} \left(\unit + \sum\limits_{n=1}^{\infty} \prod\limits_{k=1}^n \frac{-\frac{1}{2} - k + 1}{k} \Amatrix^n \right) \nonumber \\
    & = & \mu^{-\frac{1}{2}} \left( \unit  - \frac{1}{2} \Amatrix + \frac{3}{8} \Amatrix \Amatrix + ... \right) \,.
    \label{eq:Oinv_Taylor1}
\end{eqnarray}
The desired derivative of $\ooverlap^{-\frac{1}{2}}$ thus reads:
\begin{eqnarray}
    \frac{\partial \ooverlap^{-\frac{1}{2}}}{\partial \displ_K} & = & \mu^{-\frac{1}{2}} 
    \Biggl( -\frac{1}{2} \frac{\partial \Amatrix}{\partial \displ_K} \Biggr. \nonumber \\
    & & \Biggl. + \frac{3}{8} \left[ \frac{\partial \Amatrix}{\partial \displ_K} \Amatrix + \Amatrix \frac{\partial \Amatrix}{\partial \displ_K} \right] + ... \Biggr)  \,.
    \label{eq:Oinv_Taylor2}
\end{eqnarray}
It is important to note that in Eq.~\eqref{eq:Oinv_Taylor2} matrices $\Amatrix$ and $\frac{\partial \Amatrix}{\partial \displ_K}$ may not in general commute (as well as matrices in higher order terms), and therefore it is necessary to keep the correct order of matrices when taking a derivative of Eq.~\eqref{eq:Oinv_Taylor1}. To do so, Eq.~\eqref{eq:Oinv_Taylor2} can be effectively implemented via a loop over Taylor terms and using a recursive formula:
\begin{equation}
   \frac{\partial (\Amatrix^n)}{\partial \displ_K} = 
   \frac{\partial (\Amatrix \Amatrix^{n-1})}{\partial \displ_K} = \frac{\partial \Amatrix}{\partial \displ_K} \Amatrix^{n-1} +
   \Amatrix \frac{\partial (\Amatrix^{n-1})}{\partial \displ_K} \,.
   \label{eq:Taylor_recursive_eq}
\end{equation}
Finally, the only starting inputs that are needed for the recursive formula~\eqref{eq:Taylor_recursive_eq} are the matrix $\Amatrix$ [see Eq.~\eqref{eq:Taylor_Amatrix}] and its first derivative
\begin{equation}
    \frac{\partial \Amatrix}{\partial \displ_K} = \frac{1}{\mu} \frac{\partial \ooverlap}{\partial \displ_K} \,,
\end{equation}
where $\frac{\partial \ooverlap}{\partial \displ_K}$ is defined in Eq.~\eqref{eq:overlap_deriv}.

In the case of periodic solids, i.e. when Bloch sums of OAO or NAO are used, it turns out that this approximate approach based on the Taylor series is not effective. The main issue is that the
Taylor series appearing in Eq.~\eqref{eq:Oinv_Taylor2} converges very slowly, and thus in practice it requires several hundreds of terms for achieving accuracy of Hubbard forces comparable to that obtained when using the exact solution described in Sec.~\ref{sec:Lyapunov}.
This is so because many off-diagonal elements of the overlap matrix $\ooverlap$ are non-zero in the basis of Bloch-summed OAO or NAO. Possibly, faster convergence of series can be achieved by inverting the order of Taylor series expansion and evaluating Bloch sums. But the correctness of these assumptions requires more detailed investigations; this is beyond the scope of the present work, and is left for future studies. However, in the case of finite systems with localized basis sets, or for periodic solids with optimally chosen functions for the Hubbard manifold~\cite{Novoselov:2015}, the overlap matrix $\ooverlap$ might be diagonally dominant, and therefore the Taylor series can be designed with fast convergence properties.

\section{Formal solution of Lyapunov equation}
\label{app:Lyap_solution}

In this Appendix we demonstrate that Eq.~\eqref{eq:dS1} represents an exact solution of Eq.~\eqref{eq:Lyapunov_eq}. This can be verified via a direct substitution of Eq.~\eqref{eq:dS1} in Eq.~\eqref{eq:Lyapunov_eq}, which gives:
\begin{eqnarray}
    \label{eq.proofdS}
    & &  \int_0^\infty d\ttt \  \Big(\InvSqrtO e^{- \ttt \, \InvSqrtO} \ \DinvO_K \ e^{- \ttt \, \InvSqrtO} \Big. \nonumber \\
    & & \hspace{1cm} \Big. + e^{- \ttt \, \InvSqrtO t} \ \DinvO_K \ e^{- \ttt \, \InvSqrtO} \InvSqrtO \Big) \nonumber \\ & = & 
     \int_0^\infty d\ttt \
    \frac{d}{d\ttt}\Big[-e^{- \ttt \, \InvSqrtO} \ \DinvO_K \ e^{- \ttt \, \InvSqrtO}\Big]
 \nonumber \\
    & = & \Umatrix \Big[ - e^{- \ttt \, \Odiag^{-\frac{1}{2}}} \ \tilde{\DinvO}_K \ e^{- \ttt \, \Odiag^{-\frac{1}{2}}} \Big]^{\infty}_0 \Umatrix^\dag \nonumber \\
    & = & \Umatrix \tilde{\DinvO}_K \Umatrix^\dag \ = \ \DinvO_K \ ,
\end{eqnarray}
where we have used the property that each exponential in Eq.~\eqref{eq.proofdS} decays to zero in the limit $t\rightarrow \infty$, provided that all eigenvalues of the matrix $\InvSqrtO$ are positive~\footnote{Actually, the necessary condition for the integral to converge is less stringent, requiring only the real part of each eigenvalue of $\InvSqrtO$ to be positive.}. From this result, we derive the closed form presented in Eq.~\eqref{eq:dS3}.

\section{Computational scaling}
\label{sec:Scaling}

In this appendix we compare the scaling (to a leading order) of the calculation of Hubbard forces using NAO and OAO, in the framework of DFT+$U$+$V$. Essentially, the goal is to compare the computational cost in evaluating the derivative of occupation matrices given by Eq.~\eqref{eq:occ_matrix_deriv_force}. Below we present the estimates based on our current implementation of Hubbard forces.

In the case of NAO, the scaling is:
\begin{equation}
    T_\mathrm{NAO} \propto 3 N_\mathrm{at} N_\mathbf{k} \left[ T_{\mathrm{US/PAW}} + T_\mathrm{NAO}^\mathrm{proj} \right] \,,
    \label{eq:Scaling_NAO}
\end{equation}
where the factor of 3 is due to the displacements of all atoms in three Cartesian directions, $N_\mathrm{at}$ is the number of all atoms in the simulation cell, and $N_\mathbf{k}$ is the number of $\mathbf{k}$ point in the irreducible wedge of the Brillouin zone. $T_{\mathrm{US/PAW}}$ is the computational cost of evaluating the US or PAW PP related term given by Eq.~\eqref{eq:force_atomic_Sderiv2}. This latter term is the same for NAO and OAO, and thus will not be analyzed further. The quantity $T_\mathrm{NAO}^\mathrm{proj}$ represents the computational cost of evaluating the second term in Eq.~\eqref{eq:force_atomic_USterm}. The leading scaling of $T_\mathrm{NAO}^\mathrm{proj}$ can be estimated as:
\begin{equation}
    T_\mathrm{NAO}^\mathrm{proj} \propto N_\mathrm{PW} N^U_\varphi N_\mathrm{bands} \,,
     \label{eq:Scaling_NAO2}
\end{equation}
where $N_\mathrm{PW}$ is the number of plane waves in the basis set (determined by the kinetic energy cutoff), $N_\mathrm{bands}$ is the number of electronic bands (KS states), and $N^U_\varphi$ is the number of states in the Hubbard manifold of each Hubbard atom (for different types of Hubbard atoms the number of states varies depending on which atomic shell is considered as a Hubbard manifold). Generally, $T_\mathrm{NAO}^\mathrm{proj} \ll T_{\mathrm{US/PAW}}$, and the total computational cost of evaluating Hubbard forces using NAO is negligible with respect to the cost of a self-consistent iterative solution of KS equations. 

In the case of OAO, the scaling is:
\begin{equation}
    T_\mathrm{OAO} \propto 3 N_\mathrm{at} N_\mathbf{k} \left[ T_{\mathrm{US/PAW}} + T_\mathrm{ortho}^\mathrm{proj} + T_\mathrm{OAO}^\mathrm{proj} \right] \,,
    \label{eq:Scaling_OAO}
\end{equation}
where $T_{\mathrm{US/PAW}}$ is the same as in Eq.~\eqref{eq:Scaling_NAO}. In Eq.~\eqref{eq:Scaling_OAO} there is a new term ($T_\mathrm{ortho}^\mathrm{proj}$) that is not present in Eq.~\eqref{eq:Scaling_NAO}, and it is related to the evaluation of the derivative of $\ooverlap^{-\frac{1}{2}}$ (see Sec.~\ref{sec:Lyapunov}). The scaling of this latter term is:
\begin{equation}
   T_\mathrm{ortho}^\mathrm{proj} \propto 2 N_\mathrm{PW} N^\mathrm{all}_\varphi N^\mathrm{all}_\mathrm{tot} + 
   T_{\mathrm{US/PAW}}^\mathrm{ortho} + 4 \left( N^\mathrm{all}_\mathrm{tot} \right)^3 ,
   \label{eq:Scaling_OAO_ortho}
\end{equation}
where $N^\mathrm{all}_\varphi$ is the number of all orthogonalized atomic states per each Hubbard atom (obviously, for different atomic types $N^\mathrm{all}_\varphi$ is different, but here we use an average value for simplicity), $N^\mathrm{all}_\mathrm{tot}$ is the total number of all orthogonalized atomic states of all atoms in the system (if all atoms are of the same type then $N^\mathrm{all}_\mathrm{tot} = N^\mathrm{all}_\varphi N_\mathrm{at}$). The first term in Eq.~\eqref{eq:Scaling_OAO_ortho} describes the computational cost of computing the first and second terms in Eq.~\eqref{eq:overlap_deriv}. The second term in Eq.~\eqref{eq:Scaling_OAO_ortho}, $T_{\mathrm{US/PAW}}^\mathrm{ortho}$, is the analog of $T_{\mathrm{US/PAW}}$, and it describes the computational cost of evaluating the last term in Eq.~\eqref{eq:overlap_deriv}. The last term in Eq.~\eqref{eq:Scaling_OAO_ortho} describes the scaling of evaluating Eq.~\eqref{eq:dS5}, which requires four sequential matrix-matrix multiplications of size $N^\mathrm{all}_\mathrm{tot} \times N^\mathrm{all}_\mathrm{tot}$ and one inexpensive linear algebra operation [division by the term in the denominator in Eq.~\eqref{eq:dS5}] which we neglected in the estimate of the cost. The general trend is the following: $4 \left( N^\mathrm{all}_\mathrm{tot} \right)^3 < T_{\mathrm{US/PAW}}^\mathrm{ortho} < 2 N_\mathrm{PW} N^\mathrm{all}_\varphi N^\mathrm{all}_\mathrm{tot}$. Finally, the last term in Eq.~\eqref{eq:Scaling_OAO}, $T_\mathrm{OAO}^\mathrm{proj}$, describes the computational cost of evaluating Eq.~\eqref{eq:OAO_deriv} and the second term in Eq.~\eqref{eq:force_atomic_USterm} (after replacing $\frac{\partial \varphi^{I}_{m_1}}{\partial \displ_K}$ by $\frac{\partial \tilde{\varphi}^{I}_{m_1}}{\partial \displ_K}$), and it can be written as:
\begin{eqnarray}
    T_\mathrm{OAO}^\mathrm{proj} & = & N^U_\mathrm{at} \left( 1 + N^V_\mathrm{neigh} \right) \nonumber \\
    & & \times N_\mathrm{PW} N^U_\varphi \left( N_\mathrm{bands} + N^\mathrm{all}_\mathrm{tot} + N^\mathrm{all}_\varphi \right) \,,
    \label{eq:Scaling_OAO_2}
\end{eqnarray}
where $N^U_\mathrm{at}$ is the number of Hubbard atoms in the system, and $N^V_\mathrm{neigh}$ is the number of neighbors for each Hubbard atom for which the intersite Hubbard $V \ne 0$. Note that in the DFT+$U$ case, $N^V_\mathrm{neigh} = 0$ and hence the scaling in Eq.~\eqref{eq:Scaling_OAO_2} is largely reduced. By comparing Eqs.~\eqref{eq:Scaling_NAO2} and \eqref{eq:Scaling_OAO_2} it is easy to see that in the latter the scaling is largely increased in particular due to the prefactor $N^U_\mathrm{at} ( 1 + N^V_\mathrm{neigh} )$. This prefactor is present due to the fact that for OAO the derivative $\frac{\partial \tilde{\varphi}^{I}_{m_1}}{\partial \displ_K}$ is different from zero even when $K \ne I$ [see the discussion after Eq.~\eqref{eq:OAO_deriv}], while for NAO the derivative $\frac{\partial \varphi^{I}_{m_1}}{\partial \displ_K}$ is different from zero only when $K = I$. Overall, the general trend in Eq.~\eqref{eq:Scaling_OAO} is the following: $T_{\mathrm{US/PAW}} < T_\mathrm{ortho}^\mathrm{proj} \ll T_\mathrm{OAO}^\mathrm{proj}$.

Therefore, from the discussions above it becomes obvious that the computational cost of Hubbard forces using OAO is much more expensive than that using NAO, i.e. $T_\mathrm{NAO} \ll T_\mathrm{OAO}$. This is not surprising, because the former requires extra computations related to the evaluation of the derivative of $\ooverlap^{-\frac{1}{2}}$, and due to the fact that there are non-zero contributions to the Hubbard force even for non-Hubbard atoms due to the presence of the overlap matrix $\ooverlap$. Despite such an increase in the computational cost when using OAO, still this method can be efficiently implemented using standardized mathematical libraries (BLAS and LAPACK~\cite{Anderson:1999}), and effectively parallelized over plane waves, $\mathbf{k}$ points, and bands.


%

\end{document}